\documentclass[a4paper,11pt]{article}

\pdfoutput=1

\usepackage{jcappub}
\usepackage{amsmath}
\usepackage{amsfonts}
\usepackage{amssymb}
\usepackage{mathtools}
\usepackage{graphics}
\usepackage{citesort}
\usepackage{graphicx}
 \usepackage{url}

\usepackage[utf8]{inputenc}

\usepackage{hyperref}

\title{Tau lepton asymmetry by sterile neutrino emission -- Moving beyond  one-zone  supernova models}

\author[a]{Anna M.~Suliga,}
\emailAdd{anna.suliga@nbi.ku.dk}
\author[a]{Irene Tamborra,}
\emailAdd{tamborra@nbi.ku.dk}
\author[b,c]{Meng-Ru Wu}
\emailAdd{mwu@gate.sinica.edu.tw}

\affiliation[a]{Niels Bohr International Academy and DARK, Niels Bohr Institute, University of Copenhagen, Blegdamsvej 17, 2100, Copenhagen, Denmark}
\affiliation[b]{Institute of Physics, Academia Sinica, Taipei, 11529, Taiwan}
\affiliation[c]{Institute of Astronomy and Astrophysics, Academia Sinica, Taipei, 10617, Taiwan}

\abstract{The mixing of active neutrinos with their  sterile counterparts with keV mass is known to have a potentially major impact on the energy loss from the supernova core. By relying on a set of three static hydrodynamical backgrounds mimicking the early accretion phase and the Kelvin-Helmoltz cooling phase of a supernova,  we develop the first self-consistent, radial- and time-dependent treatment of $\nu_s$--$\nu_\tau$ mixing  in the dense stellar core. We follow the flavor evolution by including ordinary matter effects, collisional production of sterile neutrinos, as well as reconversions of sterile states into active ones. The dynamical feedback of the sterile neutrino production on the matter background  leads to the development of a $\nu_\tau$--$\bar\nu_\tau$ asymmetry ($Y_{\nu_\tau}$) that grows in time until it reaches a value larger than $0.15$. Our results hint towards significant implications for the supernova physics, and call for a self-consistent modeling of the sterile neutrino transport in the supernova core  to constrain the  mixing parameters of sterile neutrinos. 
}
\definecolor{green}{rgb}{0,0.5,0}
\begin{document}

\maketitle

\section{Introduction}\label{sec:introduction}
Core-collapse supernovae (SNe) are amongst the most energetic and yet mysterious transients occurring in our Universe. They originate from the death of  stars with mass larger than $8\,M_\odot$. According to our current understanding,   neutrinos play a pivotal role in the SN  mechanism by reviving the stalled shock wave and driving the SN nucleosynthesis~\cite{Mirizzi:2015eza,Janka:2012wk,Janka:2017vcp,Kotake:2012nd,Burrows:2012ew}. 

Despite the well-known importance of neutrinos in the core collapse, their behavior in the dense stellar core is not yet fully understood~\cite{Mirizzi:2015eza,Chakraborty:2016yeg}. Most importantly,  current hydrodynamical SN  simulations are run under the assumption that neutrino flavor conversions do not play any major role in the explosion mechanism~\cite{Dasgupta:2011jf,OConnor:2018sti,Janka:2016fox}. Recent findings shed doubt on this ansatz, and an agreement remains to be found~\cite{Sawyer:2015dsa,Sawyer:2008zs,Izaguirre:2016gsx,Abbar:2018beu,Dasgupta:2016dbv,Capozzi:2018clo,Richers:2019grc,Shalgar:2019kzy,Azari:2019jvr}. 

Even more fundamentally, if particles beyond the   Standard Model ones  exist, they may  dramatically affect the SN explosion dynamics. In this context,  the putative existence of  extra neutrino families, such as sterile neutrinos, may have a major impact;  interestingly, the existence of sterile neutrinos is compatible with current data coming from astrophysical, terrestrial, and cosmological surveys. See Refs.~\cite{Merle:2017dhf,Boyarsky:2018tvu,Abazajian:2012ys,Abazajian:2019ejt} for recent reviews on the topic. 

Over the years, several experiments reported evidence of sterile neutrinos in various mass ranges~\cite{Abazajian:2012ys,Boyarsky:2018tvu}. Although even eV-mass sterile neutrinos may have a non-negligible impact on the SN nucleosynthesis and dynamics~\cite{Xiong:2019nvw,Pllumbi:2014saa,Tamborra:2011is,Wu:2013gxa,Nunokawa:1997ct}, in this work we  focus on sterile neutrinos with mass between 1 and 100~keV. Sterile neutrinos with mass up to $\mathcal{O}(50)$~keV are potential dark matter candidates~\cite{Ng:2019gch,Boyarsky:2018tvu,Merle:2017dhf}.  

If heavy sterile neutrinos exist in SNe, they could have an impact on the SN physics~\cite{Shi:1993ee,Nunokawa:1997ct,Hidaka:2007se,Hidaka:2006sg,Arguelles:2016uwb,Warren:2016slz,Raffelt:2011nc}.  At the same time, the observation of neutrinos from the SN1987A   can provide bounds on the unknown mixing parameters of the sterile particles.  However, in order to derive robust constraints on  keV-mass sterile neutrinos in agreement with observations, a self-consistent treatment of the sterile neutrino transport in the SN core is necessary. The latter has not been developed yet, given the major challenges induced by the modeling of the neutrino conversion physics in the extremely dense and degenerate SN core.

In this work, for the first time, we attempt to explore  the effects of an extra sterile neutrino family ($\nu_s$) with keV mass on the transport of energy and $\tau$ neutrino lepton number in the SN core self-consistently. To this purpose, we take into account the neutrino interaction rates and carefully track the flavor evolution within the SN core by including ordinary matter effects, collisional production of sterile neutrinos, as well as reconversion effects. We  then explore the  dynamical feedback that the production of sterile particles induces on the static SN matter background. 
Our aim is to explore whether a  lepton asymmetry develops as a consequence of the active-sterile neutrino conversions. 

The paper is organized as follows. Section~\ref{sec:nu_signal} provides a general overview on the reference SN model adopted in this work. In Sec.~\ref{sec:propagation}, we model the sterile neutrino production and propagation in the stellar core without feedback effects. The generation of a $\nu_\tau$--$\bar{\nu}_\tau$ lepton number,  and related chemical potential, due to the sterile neutrino production are explored in Sec.~\ref{sec:Y_tau} together with the  corresponding feedback on the SN physics. The impact that the previously neglected physics  has on the lepton asymmetry growth is also outlined. An outlook on our findings is presented in Sec.~\ref{sec:conclusions}.  The neutrino interaction rates in the SN core, including Pauli blocking effects, are discussed in Appendix~\ref{sec:appA}. A derivation of the collisional production of sterile neutrinos  in the dense SN core taking into account reconversions effects is presented in~Appendix~\ref{sec:AppB}.

\section{Reference neutrino signal from a core-collapse supernova}\label{sec:nu_signal}

In order to model the sterile neutrino production and propagation in the SN core, we rely on the inputs of a one-dimensional  hydrodynamical simulation  of a  SN model with mass of $18.6\,M_\odot$ and  SFHo nuclear equation of state and gravitational mass of $1.4\,M_\odot$~\cite{Bollig2016}. Given the challenges involved in the modeling of the neutrino flavor evolution under extreme conditions,  we select three time snapshots from the  $18.6\,M_\odot$  model (corresponding to  post-bounce times: $t_{\mathrm{pb}} = 0.05$, $0.5$, and $1$~s), representative of the  accretion and the early  Kelvin-Helmoltz cooling SN phases. In the following, we will assume the inputs from these three selected snapshots as static hydrodynamical backgrounds and explore the dynamical implications of the production of sterile particles in the SN core. Note that this is obviously an approximation since an eventual growth of the neutrino lepton asymmetry would affect the SN evolution itself. However, our work aims at gauging the impact that the sterile neutrino production  has on the SN neutrino transport and dynamics. This SN model  has been selected as  a benchmark case;  the results presented in Sec.~\ref{sec:Y_tau}  should be qualitatively similar for any SN model. 

Deep in the SN core, the electron (anti-)neutrinos are highly degenerate and neutrinos are in thermal equilibrium with the SN medium. Therefore, the flavor-dependent (anti-)neutrino number densities are well described by a Fermi-Dirac distribution as a function of the neutrino energy $E$:
\begin{equation}
\label{eq:F-D}
\frac{dn_{\nu}}{dE} \propto \frac{1}{2 \pi^2} \frac{E^2}{e^{(E-\mu_{\nu})/T} + 1}\ ,
\end{equation}
where  $\mu_{\nu}$ is the flavor-dependent neutrino chemical potential, such that $\mu_{\nu_\mu,\nu_\tau} = \mu_{\bar{\nu}_\mu,\bar{\nu}_\tau} = 0$, and $T$ is the temperature of the SN matter. Unless otherwise specified, hereafter we will use $\hbar = c = 1$. The neutrino distribution is normalized such that  $\int dE dn_{\nu}/dE = n_{\nu}$, with $n_{\nu}$ the local flavor-dependent neutrino number density.
\begin{figure}
\centering
\includegraphics[width=6in]{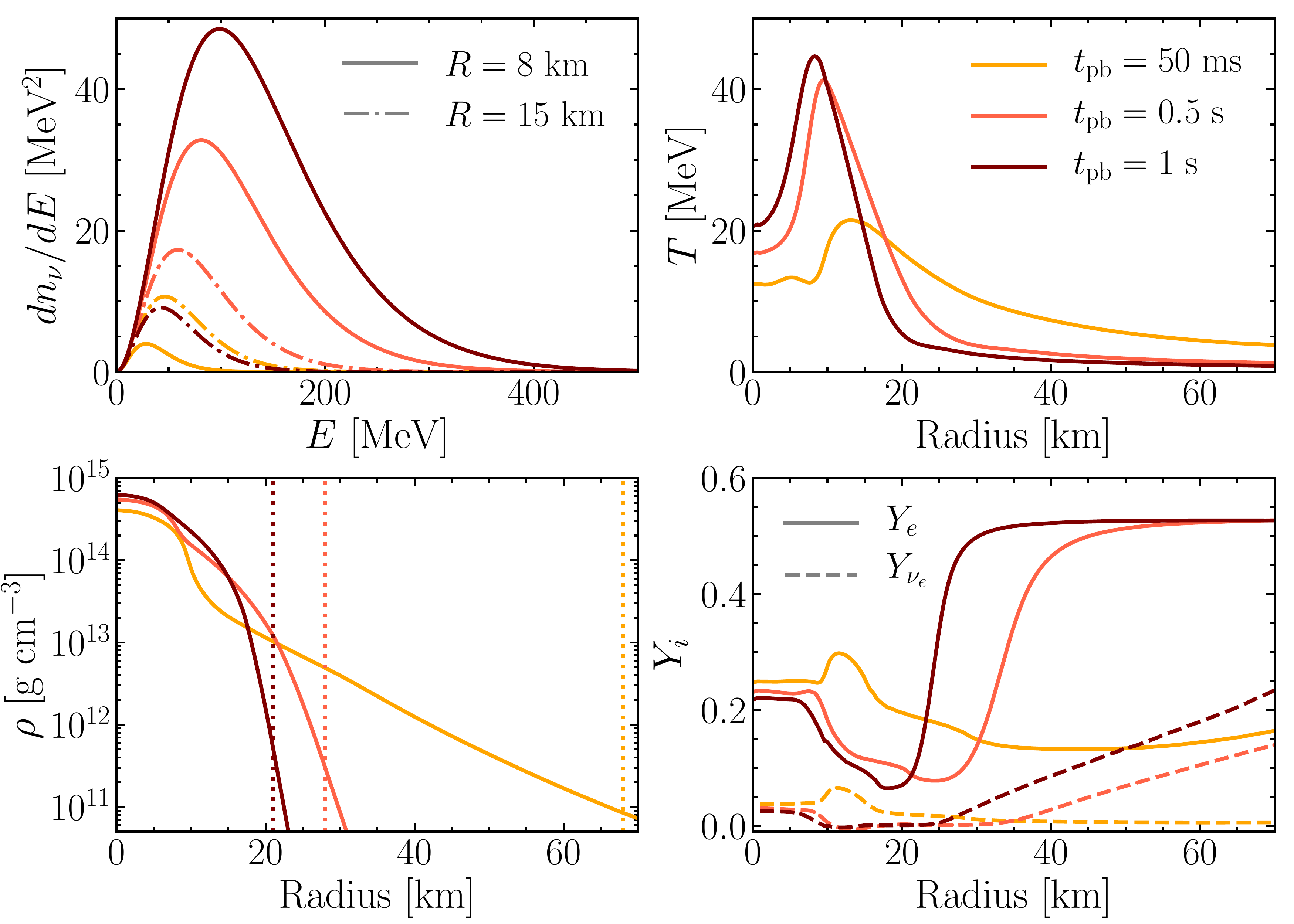}
\caption{{\it Top left}: Energy distribution of $\nu_\tau$ ($\bar\nu_\tau$) as a function of the neutrino energy  for  three selected post-bounce times ($t_{\mathrm{pb}} = 0.05$, $0.5$ and $1$~s, in yellow, pink and brown respectively) for the $18.6\,M_\odot$ SN model adopted in this work. For each time snapshot, the $\nu_\tau$ distribution is plotted at $8$ (continuous line) and $15$~km (dash-dotted) from the stellar core, respectively. Since in the absence of active--sterile flavor conversions, $\mu_{\nu_\tau} = \mu_{\bar{\nu}_\tau} = 0$, the energy distributions of $\nu_\tau$ and $\bar{\nu}_\tau$ are identical in the absence of neutrino flavor conversions. {\it Top right:} Radial profile of the medium temperature. {\it Bottom left:} Baryon  density as function of the radius; the dotted vertical lines mark the   $\nu_\tau$ neutrinosphere radius for each $t_{\mathrm{pb}}$. {\it Bottom right:} Electron fraction (solid lines) and  electron neutrino fraction (dashed lines) as a function of the radius.}
\label{fig:Inu1}
\end{figure}

The top left panel of Fig.~\ref{fig:Inu1} shows the energy distribution of $\nu_\tau$ (identical to the one of $\bar{\nu}_\tau$) in the SN core for the three selected post-bounce times: $t_{\mathrm{pb}} = 0.05$, $0.5$, and $1$~s. For each $t_{\mathrm{pb}}$, the neutrino distribution is plotted at $8$ and $15$~km, respectively.  One can see that the typical neutrino energies are $\mathcal{O}(50-200)$~MeV in the SN core. Moreover, for $t_{\mathrm{pb}} = 0.05~\text{s}$, $dn_\nu/dE$ at $15$~km is larger than the one at $8$~km. However, this does not occur for $t_{\mathrm{pb}} = 0.5$ and 1~s,  due to the proto-neutron star contraction, see top right panel of Fig.~\ref{fig:Inu1}.  For completeness, the bottom panels of Fig.~\ref{fig:Inu1} show the radial profile of the baryon density on the left, and the electron fraction ($Y_e$) together with the electron neutrino fraction ($Y_{\nu_e}$) on the right.

When (anti-)neutrinos decouple, their flavor-dependent energy distribution function deviates from a pure Fermi-Dirac distribution and it becomes pinched. It is then better represented by the following fit~\cite{Keil:2002in,Tamborra:2012ac}:
\begin{equation}
\label{eq:alpha}
\frac{dn_{\nu}}{dE} \propto  \left( \frac{E}{ \langle E_{\nu} \rangle} \right)^\beta {e^{-\frac{E (1+ \beta)}{ \langle E_{\nu} \rangle}}}\ ,
\end{equation}
where $\langle E_{\nu} \rangle$ is the first energy moment, and  $\beta$ is the pinching parameter
\begin{equation}
\beta = \frac{ \langle E_{\nu}^2 \rangle -  2 \langle E_{\nu} \rangle^2 }{\langle E_{\nu} \rangle^2 - \langle E_{\nu}^2 \rangle}\ ,
\end{equation}
depending on the second energy moment $\langle E_{\nu}^2 \rangle$. The normalization of Eq.~\ref{eq:alpha} is  such that  $\int dE dn_{\nu}/dE = n_{\nu}$.

In the following, we use Eq.~\ref{eq:F-D} (Eq.~\ref{eq:alpha}) to model the neutrino distribution before (after) the decoupling of neutrinos from the SN matter in the absence of flavor conversions. The decoupling radius  has been defined by estimating the neutrino optical depth of $\tau$ neutrinos
\begin{equation}\label{eq:optidepth}
\tau_\nu(E,r) = \int_r^\infty \sigma(E, z^\prime) n(z^\prime) dz^\prime\ , 
\end{equation}
and extracting the radius $R_\nu$ where $\tau_\nu(E,r) \simeq 1/3$. In Eq.~\ref{eq:optidepth}, 
$\sigma(E,z^\prime)$ is the cross section in the neutrino trapping region and $n(z^\prime)$ is the number density of targets (see Appendix~\ref{sec:appA}). 

Another useful quantity is  the neutrino mean free path, defining the average distance that a neutrino with energy $E$ can travel before interacting with other particles:
\begin{equation}
\lambda_\nu(E,r) \simeq \frac{1}{n(r) \sigma(E, r)}\ .
\end{equation}

Notably,  as shown in Appendix~\ref{sec:appA}, among the several processes involving neutrinos in the SN core, the scattering of neutrinos on nucleons ($\nu + N \rightarrow \nu + N$) is the dominant process (i.e., it dominates the interaction rate $\Gamma_\nu \simeq 1/\lambda_\nu$). 

\begin{figure}
\centering
\includegraphics[width=5in]{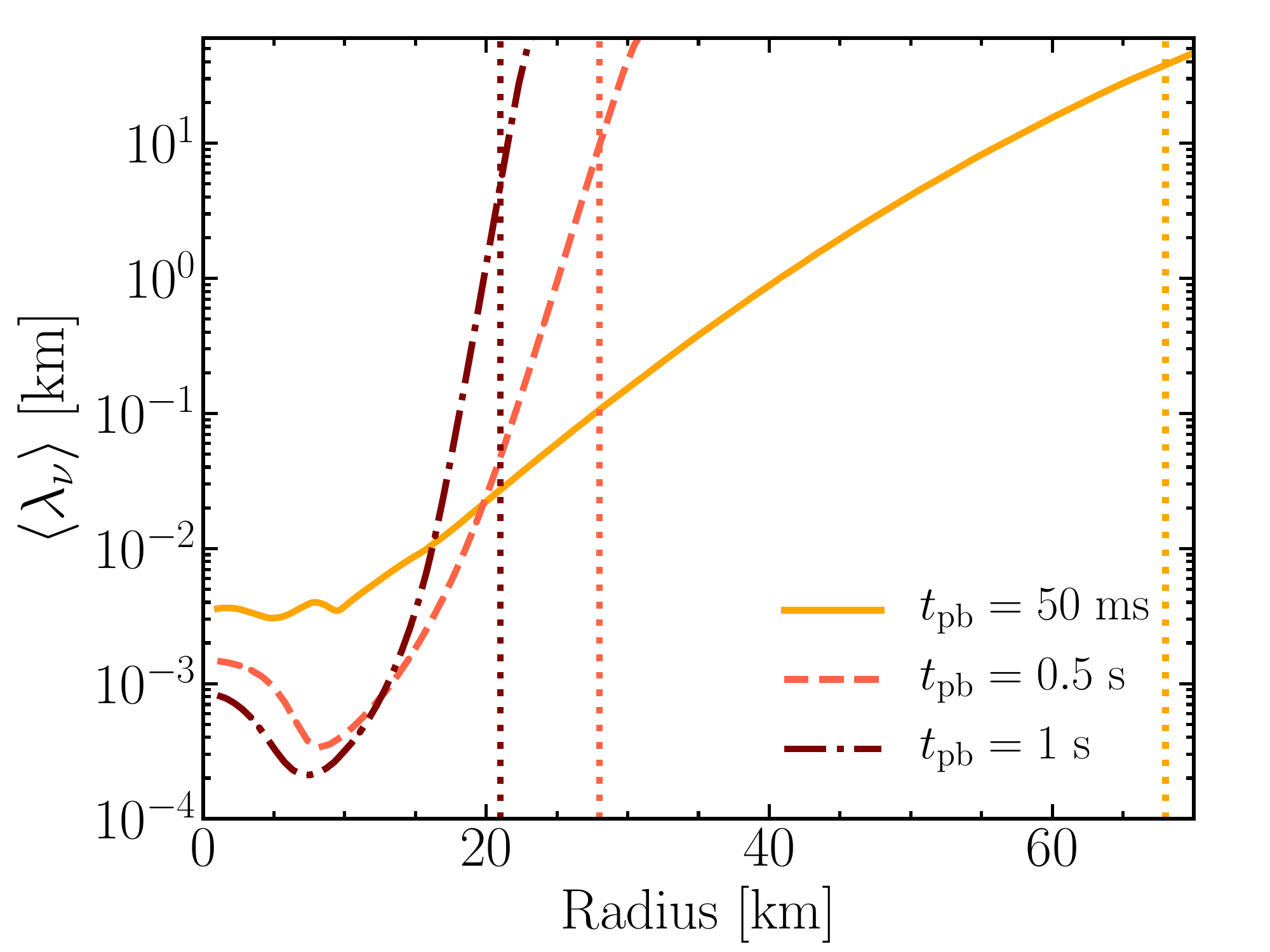}
\caption{Energy averaged mean free path of $\bar\nu_\tau$  as a function of the radius for  $t_{\mathrm{pb}} = 0.05, 0.5$ and $1$~s. Neutrinos are coupled to the matter in the SN core,  until they reach the decoupling region roughly marked by the vertical lines indicating the $\nu_\tau$ neutrinosphere radius. After decoupling, $\nu_\tau$ ($\bar\nu_\tau$)'s stream freely.}
\label{fig:diffusion}
\end{figure}
The mean free path of $\bar\nu_\tau$ averaged over the energy distribution is plotted as a function of the radius in Fig.~\ref{fig:diffusion} for $t_{\mathrm{pb}} = 0.05, 0.5$ and $1$~s. Although the mean-free path is strictly depending on the neutrino energy, one can see that the average mean free path of active (anti-)neutrinos is smaller in the SN core. On average, active neutrinos are  trapped until the decoupling region (roughly marked by the neutrinosphere radius in the plot), when  $\lambda_\nu$ becomes large enough to allow neutrinos to stream freely. The minimum of $\langle \lambda_\nu \rangle$ around $5$--$10$~km  is caused by the temperature maximum at this radius (see Fig.~\ref{fig:Inu1}).

\section{Neutrino flavor conversions in the dense stellar core}\label{sec:propagation}

In this Section, we focus on the flavor conversions in the $\tau$--$s$ sector. We  discuss matter effects leading to an enhanced production of sterile neutrinos, as well as collisional production of sterile neutrinos in the stellar core.

\subsection{Resonant production of sterile antineutrinos}\label{sec:MSW}
For the sake of simplicity, we work in a simplified two-flavor basis  $(\nu_\tau,\nu_s)$. Hence, we assume there is no mixing of the sterile neutrino eigenstate with the electron and muon neutrino. We also neglect the mixing of the active neutrinos among themselves. This is, of course, an approximation, but it should lead to reliable results for what concerns the impact of the dynamical effects due to the sterile neutrino production on the hydrodynamical quantities.  In fact, flavor conversions among the active states are expected to be suppressed in the region where active-sterile conversions occur due to the high matter density and frequent collisions with the SN medium.

The radial flavor evolution of the neutrino field is described by the Liouville equation for each energy mode $E$
\begin{equation}
\label{eq:denisty_matrix}
{\partial_r} \rho_E = {-i} [H_E, \rho_E] + \mathcal{C}(\rho_E,\bar\rho_E)\  \mathrm{and}\  \partial_r \overline\rho_E = {-i} [\overline{H}_E, \overline{\rho}_E] + \mathcal{C}(\rho_E,\bar\rho_E)\ ,
\end{equation}
where the bar denotes antineutrinos and $\rho$ is the neutrino density matrix. The density matrix for each energy mode $E$, $\rho_E$, is a  $2\times 2$ matrix in the flavor space spanned by $(\nu_\tau, \nu_s)$, and similarly for antineutrinos. The initial conditions for the neutrino field are assumed to be  $\rho_E=\mathrm{diag}(n_{\nu_\tau}, 0)$ and $\overline{\rho}_E=\mathrm{diag}(n_{\bar{\nu}_\tau}, 0)$, i.e.~we work under the assumption that $\nu_s$ are only generated through mixing with $\nu_\tau$. The Hamiltonian $H_E$  in the flavor basis takes the form
\begin{equation}
\label{hamiltonian}
H _E= H_{\mathrm{vac}, E} + H_{\mathrm{m}, E} = \frac{\Delta m_s^2}{2 E} 
\begin{bmatrix}
   -\cos 2\theta & \sin 2\theta \\
    \sin 2\theta & \cos 2\theta 
\end{bmatrix}
 + \begin{bmatrix}
    V_{\mathrm{eff}} & 0 \\
    0 & -V_{\mathrm{eff}}
\end{bmatrix}\ ,
\end{equation}
where $H_{\mathrm{vac}}$ is the vacuum term, function of the active-sterile mixing angle $\theta$ and the active-sterile mass difference $\Delta m_s^2$. The effective potential $V_{\mathrm{eff}}$ takes into account the forward scattering potential~\cite{Raffelt:2011nc}
\begin{equation}
\label{eq:potential}
V_{\mathrm{eff}} = \sqrt{2} G_F n_B \left[-\frac{1}{2}Y_n + Y_{\nu_e} + Y_{\nu_\mu} + 2 Y_{\nu_\tau}\right]\ ,
\end{equation}
with $G_F$ being the Fermi constant and $Y_i = (n_i - n_{\bar{i}})/n_B$, with $n_i$ and $n_{\bar{i}}$  the number densities of the particle species $i$ and its antiparticle $\bar{i}$. Since the distributions of $\nu_\mu$ and $\bar{\nu}_\mu$ are identical and we neglect flavor conversions in the active sector,  $Y_{\nu_\mu}=0$ initially; however,  $Y_{\nu_\tau}$  may become non-zero as sterile neutrinos are generated, as it will be discussed in Sec.~\ref{sec:Y_tau}. Because of charge neutrality $Y_p = Y_e  = 1 - Y_n$. 
Note also that the vacuum term for antineutrinos has opposite sign with respect to the one of neutrinos. 
The Mikheyev-Smirnov-Wolfenstein (MSW) resonance occurs at a certain distance $r_\mathrm{res}$ from the SN core,  when the following condition is satisfied~\cite{Mikheev:1986if,1985YaFiz..42.1441M,1978PhRvD..17.2369W}:
\begin{equation}
\cos 2\theta = \frac{2 V_{\mathrm{eff}} E_{\mathrm{res}}}{\Delta m_s^2}\ ,
\end{equation}
making the effective mixing angle in matter maximal.  

\begin{figure}
\centering
\includegraphics[width=5in]{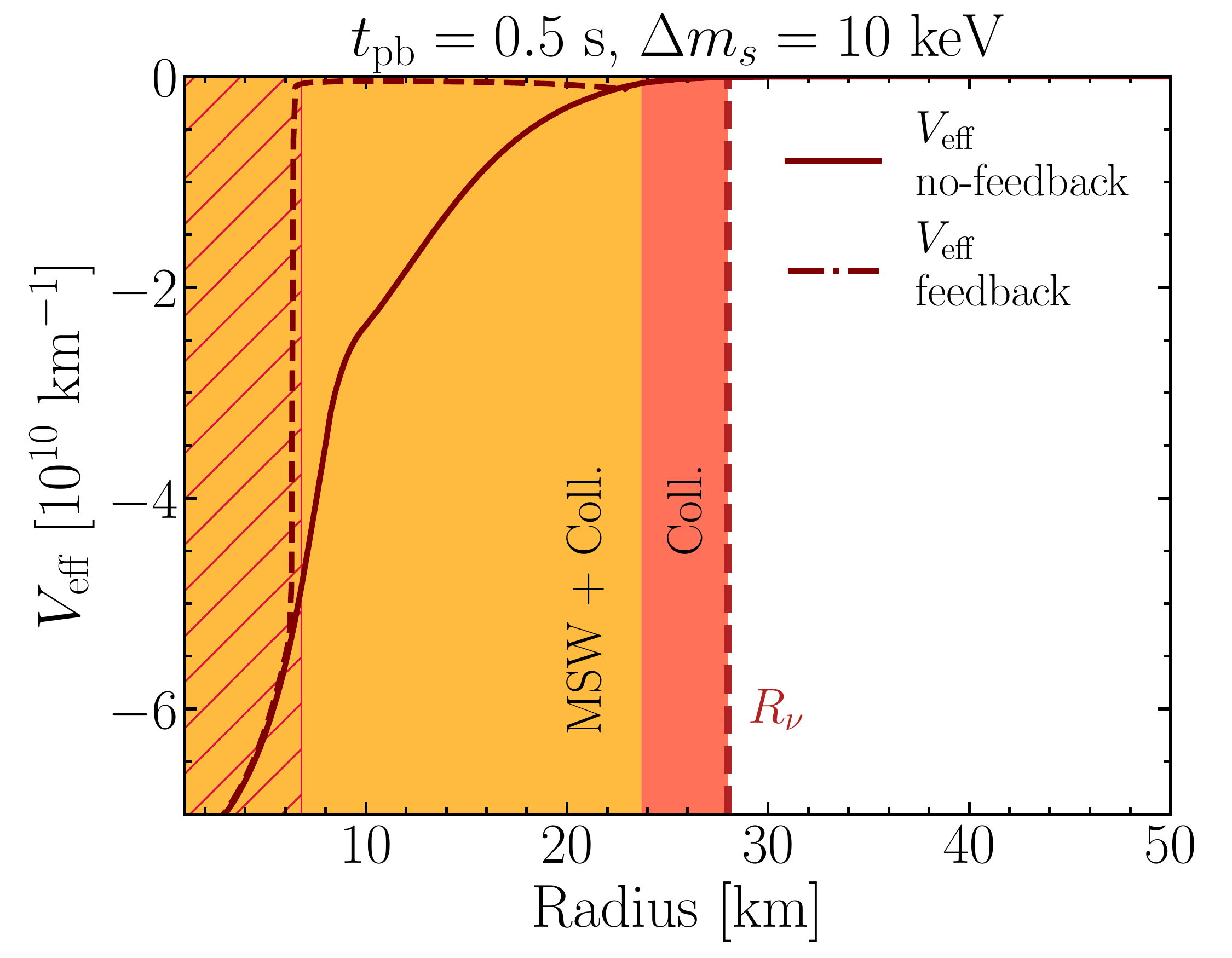}
\caption{Neutrino-matter forward scattering potential $(V_{\mathrm{eff}})$ for $\nu_\tau$--$\nu_s$ mixing as a function of the radius for $t_\mathrm{pb} = 0.5$~s with (dashed, $t_\mathrm{pb}+0.03$~s, see Sec.~\ref{sec:Y_tau} for details) feedback due to flavor conversions and in the absence of active-sterile flavor conversions (solid) . The colored bands roughly indicate the   region where MSW conversions are collisions are at play in yellow and the region where $\nu_s$ are produced through collisions in pink (hatched and plain bands are for the cases with and without feedback, respectively). The width of the band represents the location of the MSW resonances for different neutrino energies in the range $[0, 400]$~MeV for $\Delta m_s = 10$~keV. The collisional production becomes negligible when neutrinos start to free stream, i.e.~in the correspondence of the $\nu_\tau$ neutrinosphere radius (the latter is plotted in Fig.~\ref{fig:diffusion}).}
\label{fig:potential}
\end{figure}
Figure~\ref{fig:potential} shows the neutrino-matter forward scattering potential $(V_{\mathrm{eff}})$ as a function of the distance from the SN core and for the representative snapshot $t_\mathrm{pb} = 0.5$~s. In order to guide the eye, the orange  band represents the location of the MSW resonance for neutrinos with energies up to 400 MeV  and for  $\Delta m_s = 10$~keV. One can immediately gauge that sterile neutrinos are expected to be abundantly produced in the SN core. Moreover,  given the sign of $V_{\mathrm{eff}}$,  the MSW resonance conditions are only obtained for $\bar{\nu}_\tau$ and not for $\nu_\tau$.

\begin{figure}
\centering
\includegraphics[width=6in]{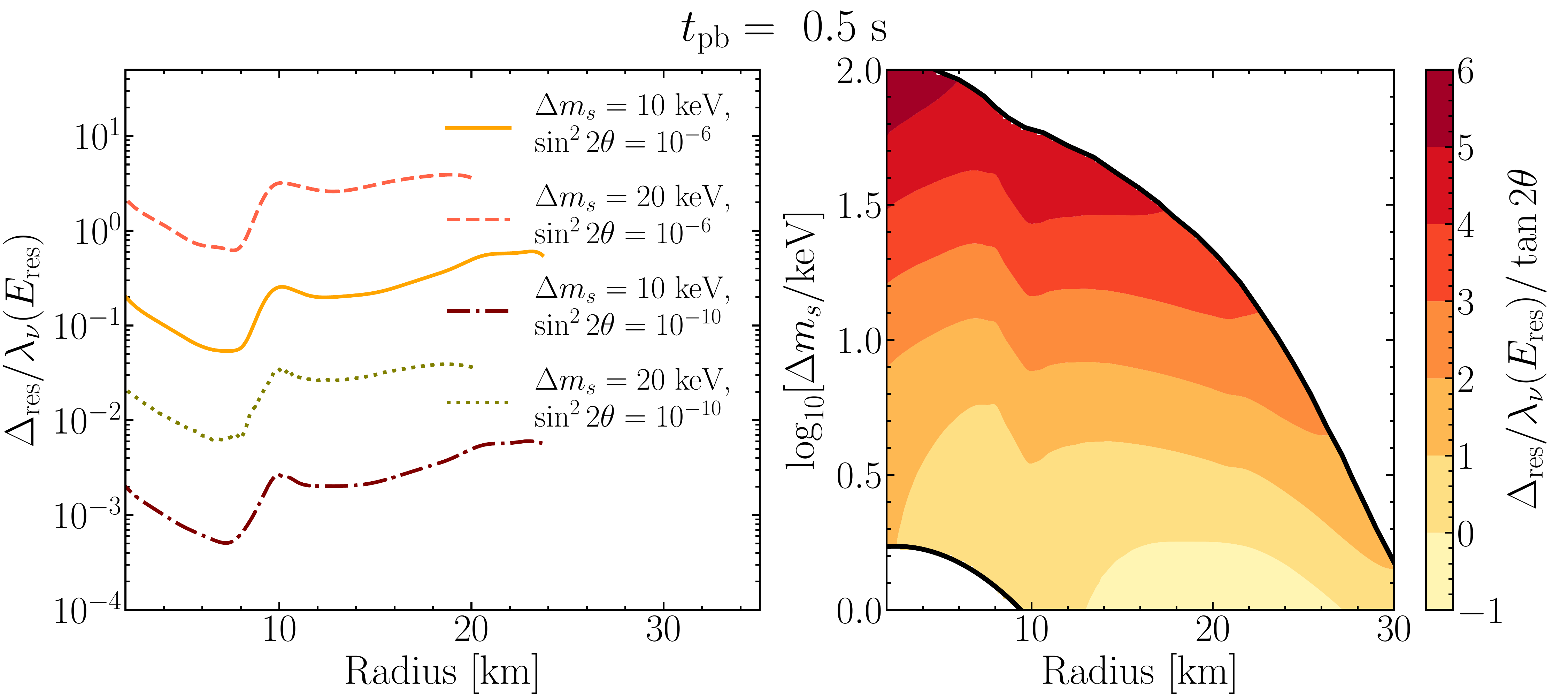}
\caption{{\it Left:} Ratio of $\Delta_{\mathrm{res}}/\lambda_{\nu}(E_\mathrm{res})$ as a function of the radius for $t_\mathrm{pb} = 0.5$~s  for different mass-mixing parameters. When $\Delta_{\mathrm{res}}/\lambda_\nu(E_\mathrm{res}) >1$, then sterile neutrinos are produced through collisions. The peak present in all curves at $\sim 10$~km is coming from the $\langle \lambda_\nu \rangle$ minimum (see Fig. \ref{fig:diffusion}). {\it Right:} Contour plot of $\Delta_{res}/(\lambda_\nu \tan 2\theta)$ in the plane defined by the radius and $\Delta m_s$ for $t_\mathrm{pb} = 0.5$~s. 
As the radius increases, $\Delta_\mathrm{res}$ decreases because more energetic neutrinos undergo MSW resonances.}
\label{fig:resonances}
\end{figure}

In the dense SN core, in order for the MSW resonance to occur, the neutrino mean free path ($\lambda_\nu$) has to be sufficiently large to allow for flavor conversions, i.e.~$\lambda_\nu \ge \Delta _{\mathrm{res}}$, with 
\begin{equation}
\Delta_{\mathrm{res}} = \tan 2\theta \left|\frac{dV_{\mathrm{eff}}/dr}{V_{\mathrm{eff}}}\right|^{-1}\ 
\end{equation} 
being the resonance width. Figure~\ref{fig:resonances} shows  $\Delta_{\mathrm{res}}/\lambda_\nu(E_\mathrm{res})$ as a function of the distance from the SN core for $t_\mathrm{pb} = 0.5$~s. One can see that $\Delta_{\mathrm{res}}/\lambda_\nu(E_\mathrm{res}) >1$ for certain neutrino mass-mixing parameters; if this happens, then sterile neutrinos can only be produced through collisions (see pink band in Fig.~\ref{fig:potential}) as we will describe in Sec.~\ref{sec:collisions}. 

When  $\lambda_\nu \ge \Delta _{\mathrm{res}}$,  the $\nu_\tau$--$\nu_s$ conversion probability is  approximated by the Landau-Zener formula~\cite{Kim:1987ss,Parke:1986jy}
\begin{equation}
\label{eq:P_MSW}
P_{\tau s}(E_{\mathrm{res}}) = 1- \exp\left({-\frac{\pi^2}{2}\gamma}\right)\ ,
\end{equation}
where $\gamma = \Delta_{\mathrm{res}}/l_{\mathrm{osc}}$, and with $l_{\mathrm{osc}} = (2 \pi E_{\mathrm{res}})/(\Delta m_s^2 \sin 2\theta)$ being the oscillation length at resonance. 

\begin{figure}
\centering
\includegraphics[width=6in]{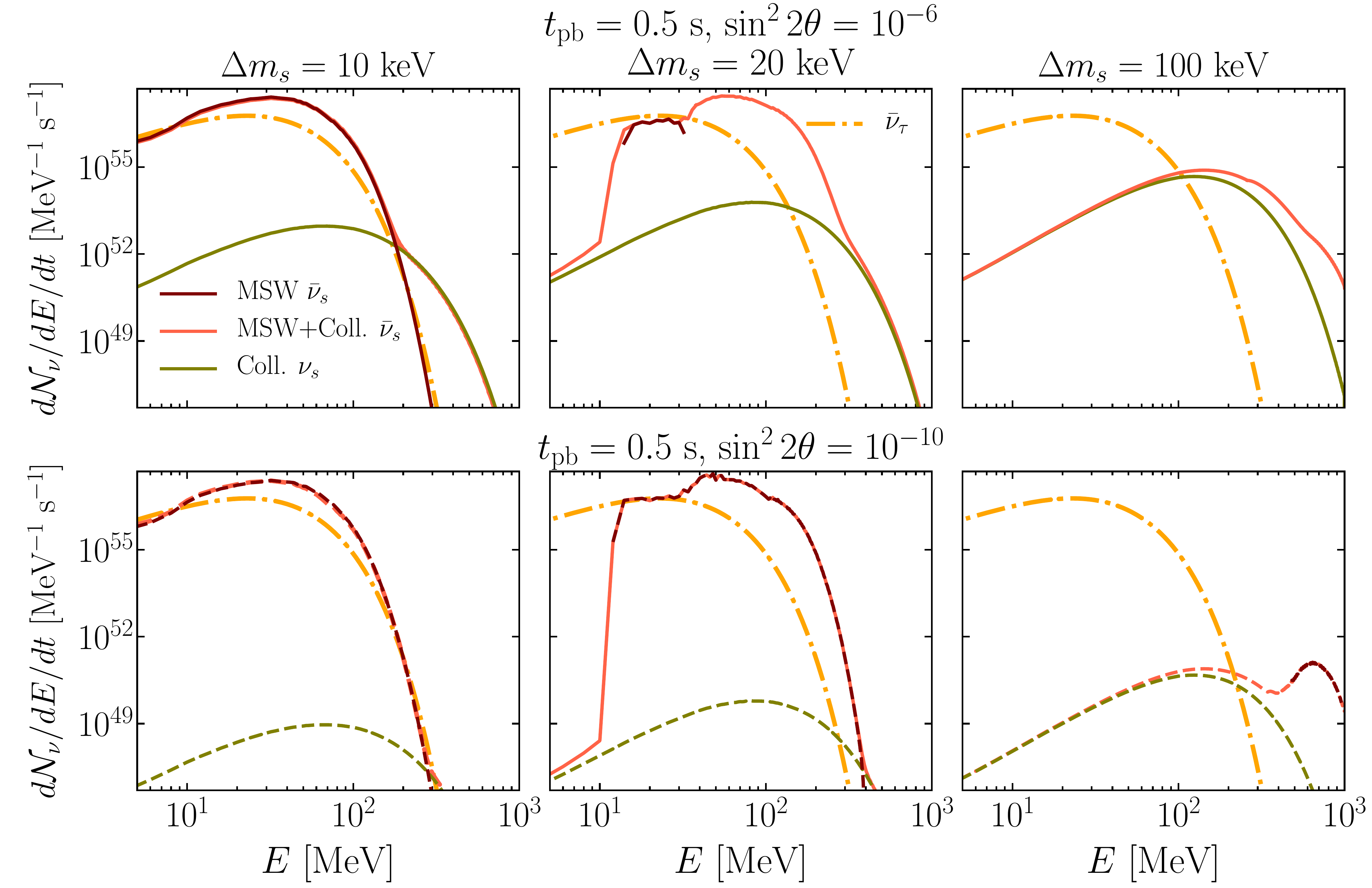}
\caption{Sterile (anti-)neutrino energy distribution as a function of the neutrino energy after MSW conversions (in~brown), MSW conversions$+$collisions (in pink) for $\bar{\nu}_{\tau}$ and only collisions for $\nu_{\tau}$ (in green) for $t_\mathrm{pb} = 0.5$~s. The mixing parameters are $\Delta m_s = 10~\mathrm{keV}$ (on the left), $\Delta m_s = 20~\mathrm{keV}$ (in the middle) and $\Delta m_s = 100~\mathrm{keV}$ (on the right), $\sin^2 2\theta = 10^{-6}$ (top panels, solid lines) and $\sin^2 2\theta = 10^{-10}$ (bottom panels, dashed lines). Additionally, the yellow dash-dotted line shows the $\bar\nu_{\tau}$ energy distribution. All distributions have been plotted at the neutrinosphere radius ($R_\nu \simeq 28$~km). Sterile particles are not produced through MSW conversions for $\sin^2 2\theta = 10^{-6}$ in the right panels (except for a tail in the high-energy part of the distribution plotted in the bottom panel) because $\Delta_{\mathrm{res}}/ \lambda_\nu > 1$. The collisional production for neutrinos and antineutrinos scales with the mixing angle due to the fact that $P_{\tau s}$ also scales with the mixing angle. Additionally in the cases where $\Delta_{\mathrm{res}}/ \lambda_\nu > 1$ the resonance width depends on the mixing angle as well.}
\label{fig:energy_spectra}
\end{figure}
In order to  track the radial evolution of the neutrino flavor, we use an adaptive grid made by multiple  shells of width $\Delta r_{\mathrm{step}, i}$ for each neutrino  energy $E_i$ in the interval $[E_{\mathrm{min}}, E_{\mathrm{max}}] = [0, 1000]$~MeV. The SN shell size is determined by $\Delta r_{\mathrm{step},i} \equiv r_i(E_i) - r_{i-1}(E_{i-1})$, where  $E_i$ and $E_{i-1}$ are the resonance energies, and  $\Delta E_i= 1$~MeV;  $\Delta r_{\mathrm{step}, i} = 0.1$~km otherwise~\footnote{Note that we adopt an energy-dependent adaptive grid; the minimum size of $\Delta r_{\mathrm{step}}$ has been chosen to be larger than the average neutrino path between collisions  and the resonant width $\Delta_\mathrm{res}$.}. For the active flavors, we assume that neutrinos are trapped for any SN shell within the neutrinosphere radius, see Fig.~\ref{fig:diffusion}. In this static scenario (without the dynamical feedback that will be implemented in Sec.~\ref{sec:Y_tau}), we assume instantaneous replenishment of the active sector.  Since in this Section we are only focusing on the radial (and not temporal) flavor evolution, sterile neutrinos are assumed to stream freely from any SN shell  because of their large mean free path. In Sec.~\ref{sec:Y_tau}, where the feedback of the sterile neutrino production on the matter background will be explored, the propagation time of the sterile states across the SN shells will be  taken into account together with the replenishment of active states.

The resulting energy distribution of sterile antineutrinos  produced through MSW resonances at the $\nu_\tau$ neutrinosphere radius is then
\begin{equation}
\label{eq:coll1}
\left(\frac{d\mathcal{N}}{dEdt}\right)_{s,\mathrm{MSW}} = \sum_{i = 1}^{N}  \Delta V_i   \frac{dn_{\nu_\tau}}{dE}(r^\prime_i)  P_{\tau s}(E_{\mathrm{res}}, r^\prime_i)\ \Delta r_{\mathrm{step},i}^{-1}\ ,
\end{equation}
where the energy-dependent differential volume of the SN shell  is $\Delta V_i = 4 \pi r_i^{\prime 2} \Delta r_{\mathrm{step},i}$ with $r^{\prime}_i$ being the central radius on which the shell of width $\Delta r_{\mathrm{step},i}$ is centered. The sum across the SN shells runs from the center of the SN core until the neutrinosphere radius $R_\nu$ (i.e., $[r_1, r_N] = [1~\mathrm{km}, R_\nu]$). Note that, since sterile neutrinos do not interact and stream freely as soon as they are produced, this is the actual number of sterile particles per unit energy and time at $R_\nu$.

Figure~\ref{fig:energy_spectra} shows the  energy distribution  for $\tau$ and sterile neutrinos and antineutrinos at $t_\mathrm{pb} = 0.5$~s after MSW conversions (in brown) for  $\Delta m_s = 10~\mathrm{keV}$ (on the left), $\Delta m_s = 20~\mathrm{keV}$ (in the middle), and $\Delta m_s = 100~\mathrm{keV}$ (on the right), $\sin^2~2\theta = 10^{-6}$ (top panels, solid lines) and $\sin^2~2\theta = 10^{-10}$ (bottom panels, dashed lines). For comparison, the yellow dash-dotted line shows the $\bar\nu_{\tau}$ energy distribution  in the absence of flavor conversions. All distributions have been plotted at the  neutrinosphere radius ($R_\nu \simeq 28$~km, see Fig.~\ref{fig:diffusion}). Given the sign of the effective matter potential, only $\bar{\nu}_s$'s are produced through MSW conversions. 
The $\bar\nu_s$ energy distributions  for $\Delta m_s~=~20~\mathrm{keV}$ are affected by MSW conversions only above $12$~MeV; in fact, lower energy modes do not undergo MSW resonances.
Sterile antineutrinos are not produced through MSW conversions for $\Delta m_s = 100~\mathrm{keV}$, except for a small fraction around $1000$~MeV because $\Delta_{res}/ \lambda_\nu > 1$, i.e.~$\bar\nu_\tau$'s interact with the matter background before to convert into sterile states. 

\begin{figure}
\includegraphics[width=6in]{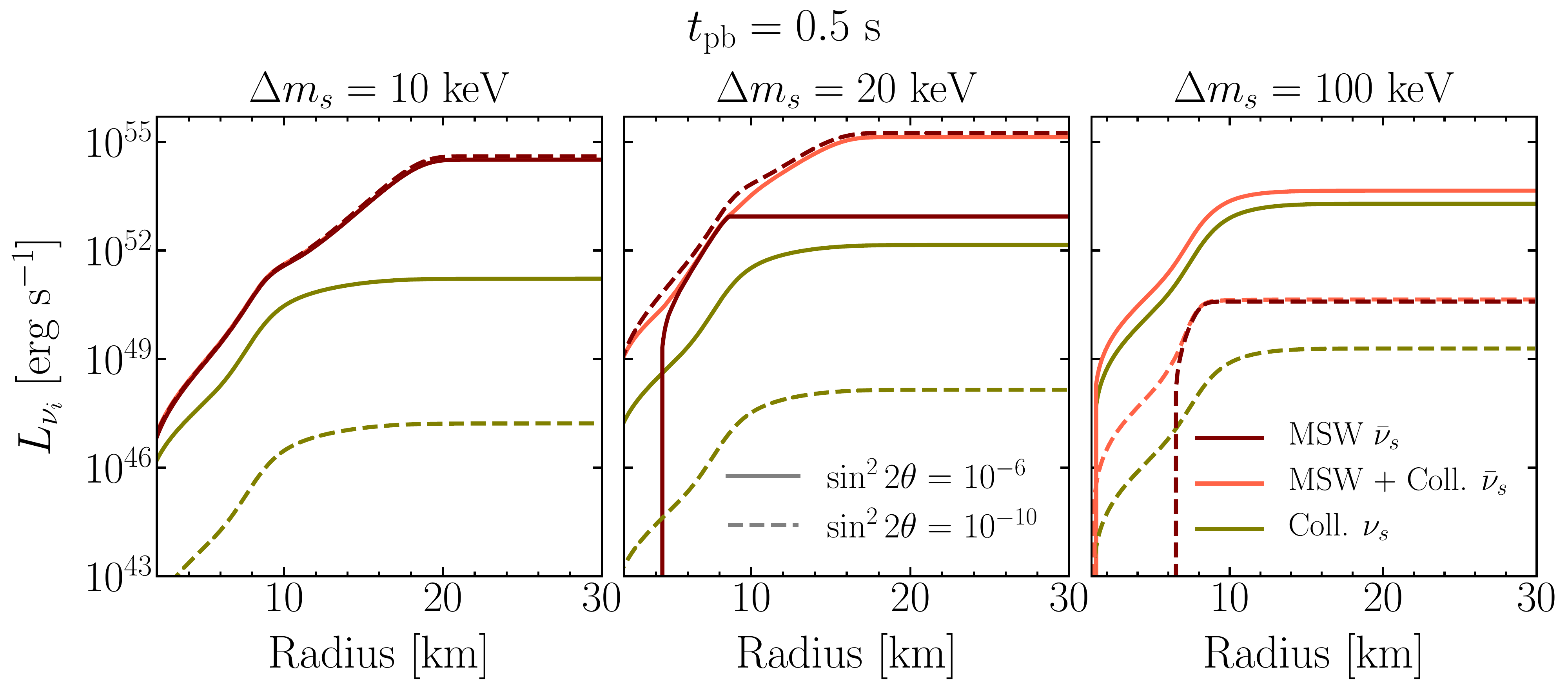}
\caption{Cumulative  luminosity of sterile neutrinos and antineutrinos as a function of the SN radius after MSW conversions (in~brown), MSW conversions$+$collisions (in pink) for antineutrinos and only collisions for neutrinos (in green) for $t_\mathrm{pb} = 0.5$~s and mixing parameters $\Delta m_s~=~10~\mathrm{keV}$ (on the left),  $\Delta m_s~=~20~\mathrm{keV}$ (in the middle), $\Delta m_s = 100~\mathrm{keV}$ (on the right), and for $\sin^2 2\theta =10^{-6}$ (solid lines) and $\sin^2 2\theta = 10^{-10}$ (dashed lines). Most of the production of sterile particles occurs within $10$--$20$~km for all the mass and mixings considered here. The collisional production of sterile particles  has a non-negligible impact as $\Delta m_s$ increases. See  Fig.~\ref{fig:energy_spectra} for comparison.}
\label{fig:number_density}
\end{figure}

The  cumulative luminosity converted in sterile antineutrinos  via MSW transitions at the distance $r$ from the SN core is
\begin{equation}
\label{eq:res}
L_{s,\mathrm{MSW}} = \sum_{j = 1}^M \sum_{k=1}^{L} \Delta E_{k} \Delta V_j  {E}_{\mathrm{res}}(r^\prime_j) \frac{dn_{\nu_\tau}}{dE_k}(r^\prime_j)  P_{\tau s}(E_{\mathrm{res}}, r^\prime_j)  \Delta r_{\mathrm{step},j}^{-1}\ ,
\end{equation}
where $[r^\prime_{j=1}, r^\prime_{j=M}] = [1~\mathrm{km}, r]$ and $[E_{k=1},E_{k=L}] = [E_{\mathrm{min}}, E_{\mathrm{max}}]$, $\Delta E_{k} = \Delta E = 1$~MeV.
Figure~\ref{fig:number_density} shows the cumulative luminosity emitted in sterile antineutrinos due to MSW conversions (in brown) for the same mass and mixing parameters shown in Fig.~\ref{fig:energy_spectra}.    As expected from Fig.~\ref{fig:potential}, MSW flavor conversions mainly occur in the SN core up to $10$--$20$~km according to the mixing.  Notably, in all cases described here
$\bar\nu_\tau$'s undergo only one MSW resonance. In fact, in the absence of collisions, the probability that  $\bar{\nu}_s$'s reconvert in $\bar\nu_\tau$'s is zero. As we will see in Sec.~\ref{sec:collisions}, reconversion effects are not negligible in the presence of collisions. In the $\Delta m_s~=~10~\mathrm{keV}$ panel, the cumulative luminosity due to MSW conversions reaches a stationary value at $\simeq 20$~km, i.e.~when MSW conversions become irrelevant. A similar behavior occurs for $\Delta m_s~=~20$ and $100~\mathrm{keV}$.

\subsection{Collisional production of sterile neutrinos and antineutrinos}\label{sec:collisions}
Given the high matter density in the SN core, active (anti-)neutrinos are trapped because of collisions, see e.g.~\cite{Raffelt:1992bs}. In the trapping regime, the neutrino propagation eigenstate is a mixture of active and sterile states;  due to  scattering on nucleons, the propagation eigenstate collapses to a pure flavor eigenstate. Sterile states can therefore  be produced through collisions. After that, the sterile states may be reconverted back to $\tau$ (anti-)neutrinos through MSW  conversions.

The effective conversion probability including the effect of collisions is 
\begin{equation}
\label{eq:Pxs_collisions}
 \langle P_{\mathrm{\tau s}}(E, r) \rangle \approx \frac{1}{4} \frac{\sin^2 2\theta}{(\cos 2\theta - 2V_{\mathrm{eff}}E/\Delta m_s^2)^2 +\sin2\theta^2 + D^2}\ ,
\end{equation}
where   $D$ is the quantum damping term,
\begin{equation}
D =\frac{ E \Gamma_{\nu}(E)}{\Delta m_s^2}\ ;
\end{equation}
 the collision rate for $\nu_\tau$ undergoing neutral current (NC) scatterings is defined as detailed in Appendix~\ref{sec:appA}:
\begin{equation}
\label{eq:nurate}
\Gamma_\nu (E) = n_B \left[F_p(E) Y_e \sigma_{\nu \mathrm{p}}(E) + F_n(E)(1- Y_e) \sigma_{\nu \mathrm{n}}(E)  \right]\ ,
\end{equation}
where $F_{p, n}(E)$ represents the Pauli blocking factor estimated as in Appendix~\ref{sec:appA}. Note that, except for the correction due to collisions, Eq.~\ref{eq:Pxs_collisions} is defined through the effective conversion probability of neutrinos in matter. 
The factor $1/4$ in Eq.~\ref{eq:Pxs_collisions} takes into account a factor $1/2$ to average the overall probability due to flavor oscillations, and another factor $1/2$ coming from taking into account the detailed balance equilibrium conditions in the SN core~\cite{Raffelt:1992bs}; note that since we include the  detailed balance correction factor in $\langle P_{\mathrm{\tau s}}(E, r) \rangle$ instead than in Eq.~\ref{eq:nurate}, Eq.~\ref{eq:Pxs_collisions} is slightly different than what has been reported in the literature, see e.g.~~\cite{Abazajian:2001nj}.  
 Equation~\ref{eq:Pxs_collisions} is such that when the damping term is large, no flavor conversion occurs and the system will be frozen in its initial state.

Figure~\ref{fig:diffusion} indicates that two scenarios are possible for active neutrinos: Either they stream freely out of the SN shell or they are trapped and undergo collisions with the dense SN matter. In the latter case, if $\lambda_\nu \ll \Delta r_{\mathrm{step}}$, neutrinos undergo multiple collisions within each SN shell of width $\Delta r_{\mathrm{step}}$. The same situation can in fact happen for sterile neutrinos.  

Once produced, according to their mean free path,  sterile neutrinos can either free stream or collide and reconvert back to active states, especially in the SN shells where favorable conditions for MSW resonances exist; unlike sterile neutrinos produced through MSW effects in the absence of collisions (see previous Section). To model this effect, we divide each SN shell of width $\Delta r_{\mathrm{step}}$ into sub-shells of width $\lambda_\nu$. The number of sub-shells within $\Delta r_{\mathrm{step}}$ is
\begin{equation}
\label{eq:n}
n = \left[ \frac{\Delta r_{\mathrm{step}}}{\lambda_\nu} \right] \ .
\end{equation}
This allows to take into account the reconversion of (anti-)$\nu_s$ to (anti-)$\nu_\tau$ within the sub-shells. The effective conversion probability has been derived in 
 Appendix~\ref{sec:AppB} and it is    
\begin{equation}
\label{eq:multiple}
P_{\tau s} (E, n) = \frac{1}{n}\sum_{k=1}^{n}   \langle P_{\tau s} \rangle (1 -  \sin^2 2\widetilde\theta \langle P_{\tau s}\rangle )^{k - 1}\ ,
\end{equation}
with $\widetilde\theta$ being the effective mixing angle in matter, under the assumption that the matter potential is constant within each SN shell (see Appendix~\ref{sec:AppB} for details). Equation~\ref{eq:multiple} indicates that the effective conversion probability is smaller as $n$ increases; moreover, for $n=1$, one recovers the case $P_{\tau s} (E, 1) =  \langle P_{\tau s} \rangle$ for $\lambda_\nu \simeq \Delta r_{\mathrm{step}}$.   Equation~\ref{eq:multiple} includes the reconversion of sterile into active states due to multiple collisions through $\langle P_{\tau s} \rangle$.

The corresponding energy distribution of sterile (anti-)neutrinos  produced through collisions until the $\nu_\tau$ neutrinosphere radius is
\begin{equation}
\label{eq:res1}
\left(\frac{d\mathcal{N}}{dEdt}\right)_{s,\mathrm{coll}} = \sum_{i=1}^{N}  \Delta V_i   \frac{dn_{\nu_\tau}}{dE}(r^\prime_i) n P_{\tau s}(E,n, r^\prime_i) \Delta r_{\mathrm{step},i}^{-1} \ .
\end{equation}
Figure~\ref{fig:energy_spectra} shows the  sterile neutrino energy distributions for the cases where  MSW resonant conversions as well as collisions are taken into account for antineutrinos  (in pink)  and  for neutrinos (collisions only, in green) for $t_{\mathrm{pb}}= 0.5$~s.  While the MSW conversions only affect the $\bar{\nu}_s$ energy distribution as discussed in Sec.~\ref{sec:MSW}, the collisional production occurs  for neutrinos and antineutrinos. As we will discuss in Sec.~\ref{sec:Y_tau}, this implies that the development of a neutrino lepton asymmetry is only possible in the presence of MSW resonances or MSW enhanced collisions. The pink line in Fig.~\ref{fig:energy_spectra} for $10~\mathrm{keV}$ and $\sin^2 2\theta = 10^{-10}$ (left, bottom panel) lies very close to the green one for energies $>~250~\mathrm{MeV}$, i.e.~the energy distribution is dominated by the collisional contribution at those high energies. In fact, antineutrinos produced through collisions before the MSW layer  reconvert back to $\bar\nu_\tau$ for fully adiabatic MSW resonances; however, this may not be the case according to the adiabaticity of MSW conversions and frequency of collisions (see $\Delta_\mathrm{res} / \lambda_{\nu}(E_\mathrm{res})$ in Fig.~\ref{fig:resonances}). As the neutrino energy increases, the MSW conversions stop being fully adiabatic for $\sin^2 2\theta = 10^{-10}$; hence, not all $\bar\nu_s$'s produced before the MSW layer reconvert to active ones. In the fully adiabatic case ($\sin^2 2\theta = 10^{-6}$, top left panel), the pink line lies slightly below the green one because of reconversions.  The collisional production dominates the high energy tail of the energy distribution also for $\Delta m_s = 20$~keV. On the other hand, at all energies, antineutrinos produced through collisions are more abundant than the ones produced through MSW conversions for $\Delta m_s = 100$~keV.

Similarly to Eq.~\ref{eq:res},  the cumulative  luminosity emitted in sterile (anti-)neutrinos  through collisional production is
\begin{equation}
\label{E_r}
L_{s, \mathrm{coll}} =  \sum_{j=1}^M \sum_{k=1}^{L} \Delta V_j \Delta E_k\ \frac{dn_{\nu_\tau}}{dE_k}(r^\prime_j)\ E_k\  \Delta r_{\mathrm{step},j}^{-1}\ n {P}_{\tau s}(E_k, n, r^\prime_j)\ ,
\end{equation}
where the numerical inputs adopted for $\Delta E$, $\Delta r_\mathrm{step}$, and the integrations over radius and energy are the same as  defined in Sec.~\ref{sec:propagation}. 
Due to the reconversion effects, the overall production of sterile particles may be suppressed according to the mixing parameters. 

Figure~\ref{fig:number_density} shows the  cumulative luminosity emitted in sterile particles, due to MSW conversions$+$collisions  for antineutrinos (in pink) and only due to collisions for neutrinos (in green), as a function of the radius for the same mass and mixing parameters shown in Fig.~\ref{fig:energy_spectra}.  

In the left panel, the MSW only case (brown curve) leads to $L_{\bar\nu_s}$ that is identical to the case of MSW conversions$+$collisions (pink curve); in fact, reconversions are only relevant for very high energies and have  a negligible impact on the overall luminosity. As also visible from Fig.~\ref{fig:potential}, collisions are important in the SN core where neutrinos are trapped and negligibly affect the sterile neutrino luminosity at radii larger than 10 km for all mass and mixing parameters shown here. In the $\Delta m_s~=~10~\mathrm{keV}$ panel, the pink line shows a bump at $\simeq 10$~km, i.e.~in correspondence of the radius where collisions stop being dominant and the medium temperature reaches its  maximum.
The collisional production of sterile particles  has a non-negligible impact as $\Delta m_s$ increases, see e.g.~the $\Delta m_s~=~100~\mathrm{keV}$  panel on the right where the particle production is dominated by collisions.   In the $\Delta m_s~=~20~\mathrm{keV}$ panel for $\sin^2 2\theta =10^{-6}$, one can see that the pink line is considerably above the green and the brown ones for $\sin^2 2\theta =10^{-6}$ because the resonant layer is collisionally enhanced  (i.e., $\Delta r_{\mathrm{res}}/\lambda_\nu > 1$, see Fig.~\ref{fig:resonances}).  In the right panel, one can see that the collisional production  is the main mechanism producing sterile particles for  $\sin^2 2\theta =10^{-6}$.

\section{Development of the neutrino lepton asymmetry and feedback effects}\label{sec:Y_tau}
In this Section, we explore the dynamical feedback that the production of sterile particles induces on the growth of $Y_{\nu_\tau}$ and, in turn, on the flavor conversion physics. 
First, we detail the treatment of the dynamical feedback on the growth of $Y_{\nu_\tau}$   in Sec.~\ref{sec:feedback}. 
Then, in Secs.~\ref{sec:4.2} and \ref{sec:parspace}, we explore  the effect of the net  $Y_{\nu_\tau}$ generation 
on the effective potential felt by neutrinos and the associated sterile neutrino production. 
In Sec.~\ref{sec:chempot}, we further discuss results derived when the feedback effects coming from the net lepton number $Y_{\nu_\tau}$, 
as well as the $\nu_\tau$ chemical potential ($\mu_{\nu_\tau}$) are taken into account.
For the sake of simplicity, we rely on a static hydrodynamical background for the three selected snapshots of the $18.6\ M_\odot$ model, and investigate the radial and temporal evolution of the neutrino lepton asymmetry and its corresponding impact on the matter potential $V_{\mathrm{eff}}$.

\subsection{Radial and temporal evolution of the neutrino lepton asymmetry}\label{sec:feedback}

Sterile neutrinos and antineutrinos are produced at the expense of $\tau$ neutrinos and antineutrinos. As discussed in Sec.~\ref{sec:propagation}, MSW (collisionally enhanced) transitions affect neutrinos and antineutrinos differently, leading to the development of $Y_{\nu_\tau}\neq 0$. One can see from Eq.~\ref{eq:potential}, and by comparing the continue and the dashed lines of Fig.~\ref{fig:potential},  that any variation of  $Y_{\nu_\tau}$ can substantially affect $V_{\mathrm{eff}}$. In turn, the modified $V_{\mathrm{eff}}$ has an impact on the flavor conversion evolution.

To investigate the dynamical feedback due to the growth of  $Y_{\nu_\tau}$ on $V_{\mathrm{eff}}$, we rely on the static hydrodynamical inputs of the three time snapshots  of the $18.6\ M_\odot$ SN model: $t_{\mathrm{pb}}=0.05, 0.5$ and $1$~s; for each time snapshot, we track the flavor evolution in $r$ as described in Sec.~\ref{sec:propagation}  simultaneously with the radial evolution of $Y_{\nu_\tau}$. The $V_{\mathrm{eff}}$ obtained by  $Y_{\nu_\tau} \neq 0$ is then adopted as input for the next temporal step. This allows to explore how $Y_{\nu_\tau}$ evolves in time $t$. Hence, although we rely on static hydrodynamical inputs for the sake of simplicity, we explore the dynamical feedback of the production of sterile particles self-consistently, by following the time and radial evolution of the relevant quantities.

For fixed $t_{\mathrm{pb}}$ and $r$, we compute the  evolution of $Y_{\nu_\tau}$ after a time interval $\Delta t$ by tracking the  flavor evolution as described in Sec.~\ref{sec:propagation}  and by solving  the following equation simultaneously
\begin{gather}
\label{eq:Y_nu_tau}
\begin{aligned}
Y_{\nu_\tau}(r, t_\mathrm{pb}+\Delta t) ={}& \frac{1}{n_B(r)} \sum_{l=1}^{P} \sum_{k=1}^{L} \left[\bar{P}_{\tau s}(E_k, t^\prime_l) \frac{dn_{\bar\nu_\tau}}{dE_k}(t^\prime_l,\Delta r_\mathrm{step}) 
 -  P_{\tau s}(E_k, t^\prime_l) \frac{dn_{\nu_\tau}}{dE_k}(t^\prime_l, \Delta r_\mathrm{step})  \right] \\ & \times \Delta E_k \Delta r_\mathrm{step}^{-1} \ \Delta t^\prime_l \ ,
\end{aligned}
\end{gather}
where $[t_{l=1}, t_{l=P}]=[t_\mathrm{pb}, t_\mathrm{pb}+\Delta t]$, $[E_{k=1},E_{k=L}] = [E_{\mathrm{min}}, E_{\mathrm{max}}]$, $\bar{P}_{\tau s}$ is the antineutrino conversion probability (Eqs.~\ref{eq:P_MSW}, \ref{eq:multiple}) and $P_{\tau s}$ is the neutrino conversion probability (Eq.~\ref{eq:multiple});  $dn_{\nu_\tau}/dE(t^\prime,\Delta r_\mathrm{step})$ is the local density of particles in $\Delta r_\mathrm{step}$, per unit energy. The radial step assumed in this calculation including the dynamical feedback is $\Delta r_\mathrm{step} \simeq 0.1~\mathrm{km}$; if $\Delta_\mathrm{res} \ge 0.1~\mathrm{km}$  for certain energies, then we assume   $\Delta r_\mathrm{step} =\Delta_\mathrm{res}$. The energy step is $\Delta E = E_\mathrm{res}(r_i+\Delta r_\mathrm{step}) - E_\mathrm{res}(r_i)$ for the MSW conversions and the MSW enhanced collisional production; instead for collisions outside the resonance, 
$\Delta E = 2~\mathrm{MeV}$. 

The time step adopted in the numerical runs is $\Delta t^\prime = 10^{-7}$~s. Sterile neutrinos produced in the trapping regime ($r < R_\nu$) take $t^\prime_\star = R_\nu/c \simeq 10^{-4}$~s $\gg \Delta t^\prime$ to escape from the neutrinosphere. Hence, for the sake of simplicity, we let all sterile particles to stream freely from the neutrinosphere every $t^\prime_\star$ (while keep them trapped in the production SN shell  for smaller times), including the reconversion effects.  In order to take into account the replenishment of $\tau$ (anti-)neutrinos through Bremsstrahlung (see Appendix~\ref{sec:appA} for details), we limit the sterile neutrino production rate by the Bremsstrahlung one if the production rate of sterile particles is larger than  $\Gamma_{NN \rightarrow NN \nu_\tau\bar\nu_\tau}$.

\subsection{Dynamical feedback due to the production of sterile particles}\label{sec:4.2}
In the following, we investigate the role of the dynamical feedback on $Y_{\nu_\tau}$ and on the sterile production, by focusing on the benchmark time snapshot $t_\mathrm{pb} = 0.5$~s for $(\Delta m_s, \sin^2 2\theta)=(10\ \mathrm{keV}, 10^{-10})$. We will generalize our findings to the three considered $t_\mathrm{pb}$ and other $(\Delta m_s, \sin^2 2\theta)$ in Sec.~\ref{sec:parspace}.
Note that, in this subsection and in Sec.~\ref{sec:parspace},
we take into account the feedback effects by updating the effective potential $V_{\rm eff}$ via Eq.~\eqref{eq:potential}, while keeping the $\nu_\tau$ and $\bar\nu_\tau$ energy distributions unaffected in Eq.~\eqref{eq:Y_nu_tau}. However, $Y_{\nu_\tau} \neq 0$ is also responsible for the growth of $\mu_{\nu_\tau} \neq 0$, which in turn would dynamically modify the (anti-)neutrino energy distributions;
results obtained by including the dynamical feedback of $Y_{\nu_\tau}$ and  $\mu_{\nu_\tau}$ are  discussed in Sec.~\ref{sec:chempot}.

\begin{figure}
\includegraphics[width=6in]{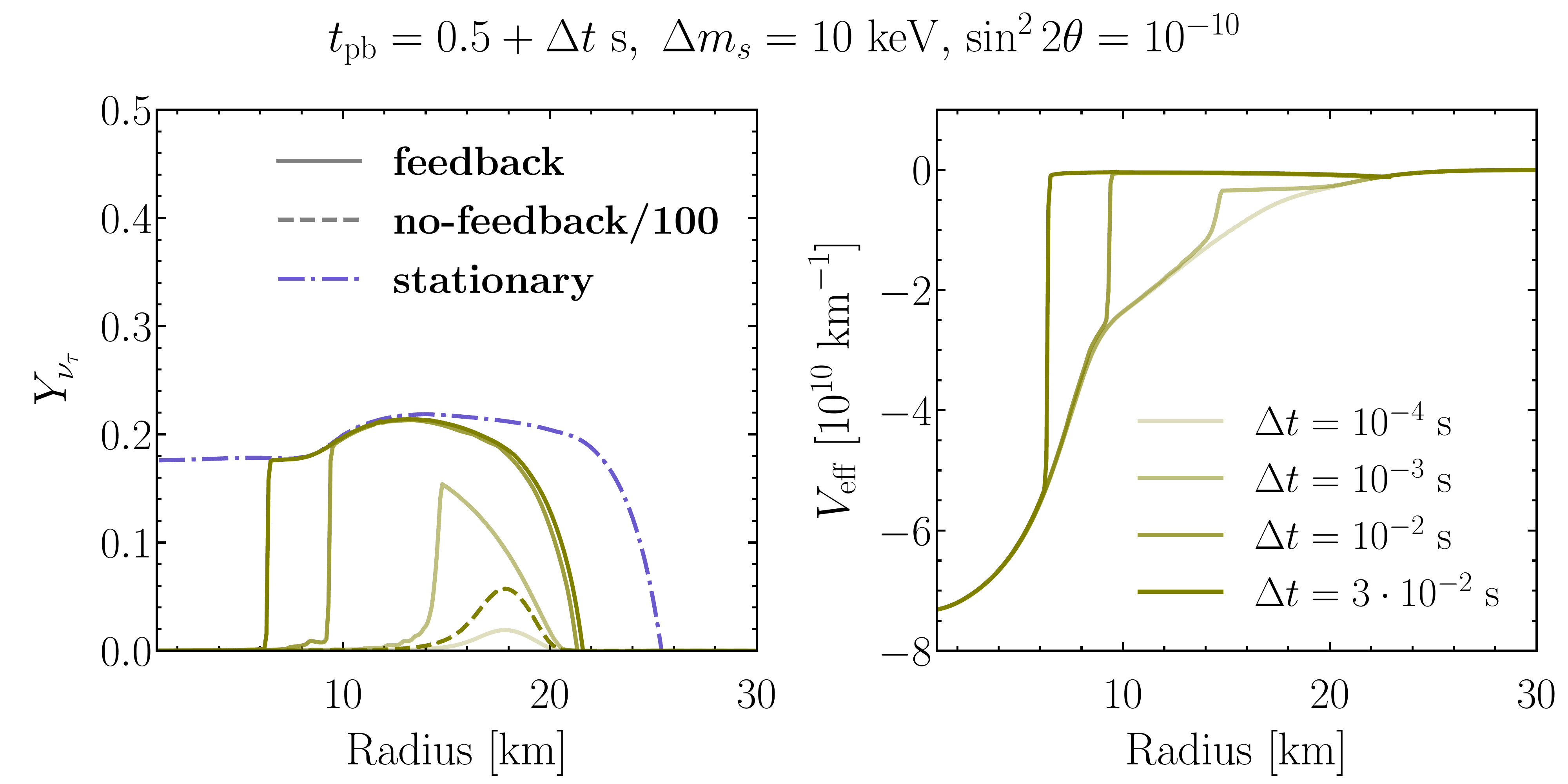}
\caption{{\it Left:} Radial profile of  $Y_{\nu_\tau}$ for $t_\mathrm{pb} = 0.5$~s, $(\Delta m_s, \sin^2~2\theta)=(10\ \mathrm{keV}, 10^{-10})$; the dashed line is for the case without feedback ($\Delta t = 0$) and continuous line for the case with feedback  due to the production of sterile particles ($\Delta t = 0.03$~s). The continuous lines with different hues  mark the temporal evolution (from lighter to darker as time increases). The dash-dotted purple curve  shows the $Y_{\nu_\tau}^{\rm stat}$  for $E_\mathrm{max} = 1000$~MeV estimated analytically (see text for details).  {\it Right:}  Corresponding radial profile of $V_{\mathrm{eff}}$ for the case  with feedback.  The  dynamical feedback due flavor conversions in sterile states considerably affects $Y_{\nu_\tau}$ and $V_{\mathrm{eff}}$. 
}
\label{fig:ytau}
\end{figure}

\begin{figure}
\includegraphics[width=6in]{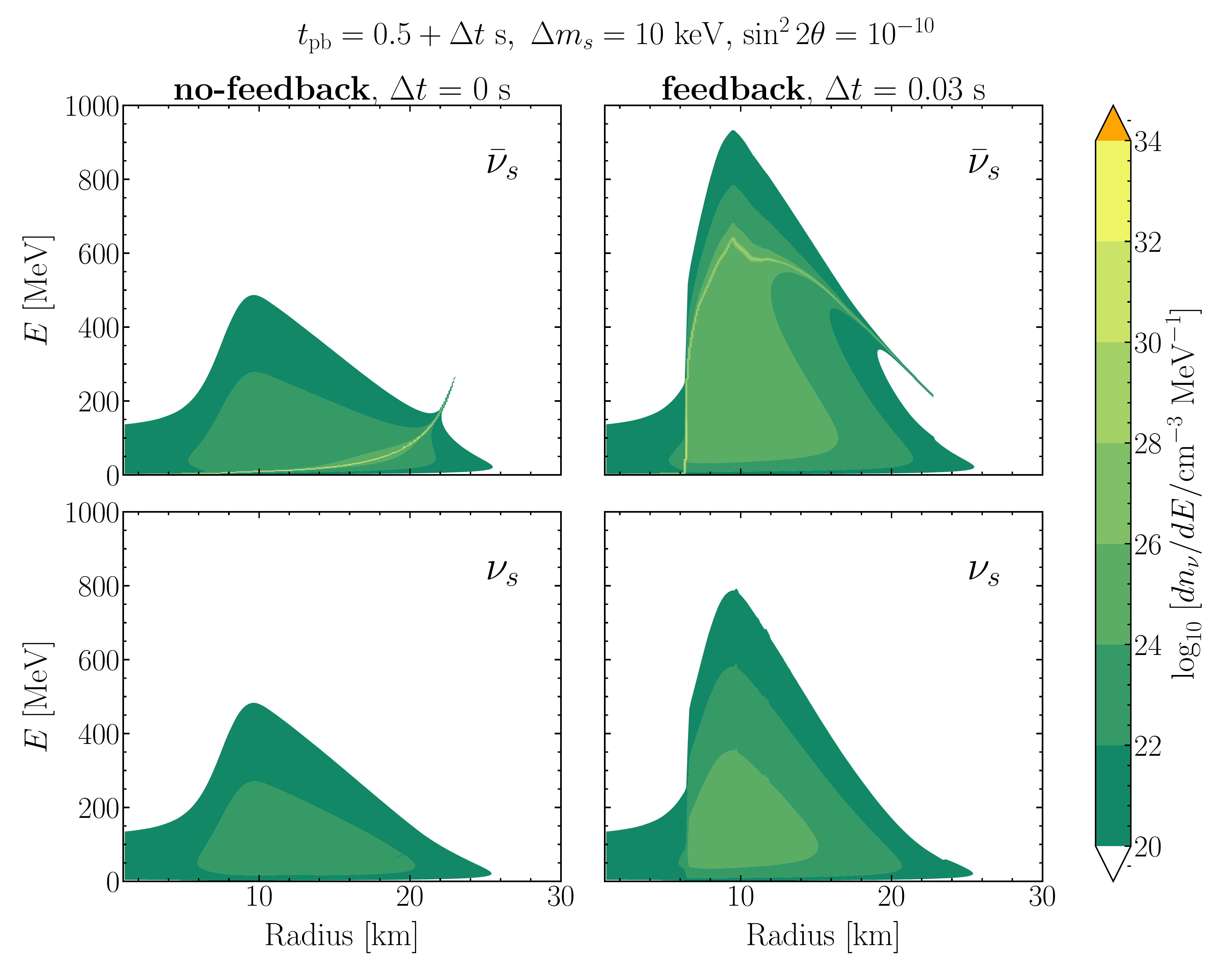}
\caption{Contour plots of the  number density of $\bar\nu_s$'s (top panels) and $\nu_s$'s (bottom panels) per unit energy in the plane spanned by the distance from the SN core and the neutrino energy  for $(\Delta m_s, \sin^2~2\theta)=(10\ \mathrm{keV}, 10^{-10})$, $t_\mathrm{pb} = 0.5$~s. {\it Left:} Without any dynamical feedback effect due to the production of sterile particles ($\Delta t = 0$), see Sec.~\ref{sec:propagation}.  {\it Right:} With dynamical feedback effect due to the production of sterile particles ($\Delta t = 0.03$~s). The dynamical modification of $V_{\mathrm{eff}}$ due to the production of sterile particles drastically affects the local density of sterile (anti-)neutrinos. The light green line in the plots on the top panels marks the location of the MSW resonances.}
\label{fig:nd}
\end{figure}

 The left panel of Fig.~\ref{fig:ytau} shows a comparison between the $Y_{\nu_\tau}$ obtained without any dynamical feedback effect due to the sterile neutrino production (dashed line), as described in Sec.~\ref{sec:propagation}, and the one obtained by tracking the radial and temporal evolution of $Y_{\nu_\tau}$ as from Eq.~\ref{eq:Y_nu_tau} (continuous line). Note that $Y_{\nu_\tau}$ has been divided by $100$ in the former case.  The continuous lines with different hues  mark the temporal evolution (from lighter to darker as time increases) within $\Delta t=0.03$~s. It shows that  $Y_{\nu_\tau}$  quickly reaches
stationary values after $\sim 0.01$~s in the region between $10$--$20$~km, and  a plateau of $Y_{\nu_\tau}\simeq 0.21$ develops.
The plateau continues to extend to smaller radii
as time progresses. We consider $\Delta t = 0.03$~s, since after such a time interval, $Y_{\nu_\tau}$ has reached  its stationary radial profile.  
As $t^\prime$ increases,  the $Y_{\nu_\tau}$  plateau  extends  towards the inner core of the SN. In fact, as $t^\prime$ increases, high-energy modes undergo MSW resonances (as we will discuss, this is also visible in the right panel of Fig.~\ref{fig:nd}). The sharp growth of $Y_{\nu_\tau}$ at small radii is due to the fact  that $Y_{\nu_\tau}$ evolves towards its stationary profile because of  MSW (collisionally enhanced) conversions, but only energy modes higher than a certain energy undergo resonances. Moreover,  the production of sterile particles becomes slower because of Pauli blocking  effects (see middle panel of Fig.~\ref{fig:Pauliblock}).  Importantly,  $Y_{\nu_\tau}$ is largely overestimated when the dynamical feedback effects are not taken into account.

At the same time, $V_{\mathrm{eff}}$ (see Eq.~\ref{eq:potential}) follows the $Y_{\nu_\tau}$ evolution, hence the radial evolution of $V_{\mathrm{eff}}$ is strongly affected, as shown in the right panel of Fig.~\ref{fig:ytau}.  The stationary value of $Y_{\nu_\tau}$ is reached when $V_{\mathrm{eff}} \rightarrow 0$ and no more conversions into sterile states occur, i.e.~$Y_{\nu_\tau}$ does not change further.  To verify this, we have analytically computed the stationary $Y_{\nu_\tau}$ value  ($Y_{\nu_\tau}^{\rm stat}$) at each radius
by equating $V_{\rm eff}$ (Eq.~\ref{eq:potential}) to $(\Delta m_s^2\cos 2\theta)/2E_{\rm max}$. The dash-dotted purple curve in Fig.~\ref{fig:ytau} shows the $Y_{\nu_\tau}^{\rm stat}$ as
a function of the radius obtained for $E_\mathrm{max} = 1000$~MeV. It clearly illustrates that, in the SN core, the sterile neutrino productions leads to the evolution of $Y_{\nu_\tau}$  towards a stationary value. We note that for the inner part ($r\lesssim 6$~km) and the outer 
part ($r\gtrsim 20$~km), it takes longer than $\sim 0.03$~s for $Y_{\nu_\tau}$ to 
reach $Y_{\nu_\tau}^{\rm stat}$, due to the slower production rate of sterile neutrinos.
As one can see from this example, the  dynamical feedback of flavor conversions considerably modifies $Y_{\nu_\tau}$ and $V_{\mathrm{eff}}$. As shown in Fig.~\ref{fig:potential}, because of the evolution of $V_{\mathrm{eff}}$ in time, neutrinos interact with a different  effective matter background, and the radial regions where MSW resonances occur evolve accordingly. 

Due to the $Y_{\nu_\tau}$ temporal evolution, the sterile (anti-)neutrino production is in turn affected. Figure~\ref{fig:nd} shows contour plots of the  number density of $\bar\nu_s$'s (on the top) and $\nu_s$'s (on the bottom) in the plane defined by the radius and  the neutrino energy  for  $\Delta t  = 0$ on the left and  $\Delta t  = 0.03$~s on the right. The  left panels represent the sterile number densities  obtained by neglecting the dynamical feedback effects due to the production of $Y_{\nu_\tau}\neq 0$ on $V_{\mathrm{eff}}$, as described in Sec.~\ref{sec:propagation}. The impact of the dynamical feedback on the sterile number density is shown in the panel on the right. One can see that a  light green line visible at the center of the contour plot for $\bar\nu_s$'s; this is due to the resonant production of sterile antiparticles. Moreover, the dynamical modification of $V_{\mathrm{eff}}$ due to the production of sterile particles drastically affects the local density of sterile antineutrinos since the temporal evolution of $V_{\mathrm{eff}}$ triggers the MSW (collisionally enhanced) production of $\bar\nu_s$'s  with higher energies (see also Fig.~\ref{fig:ytau}). 

\begin{figure}
\centering
\includegraphics[width=4.5in]{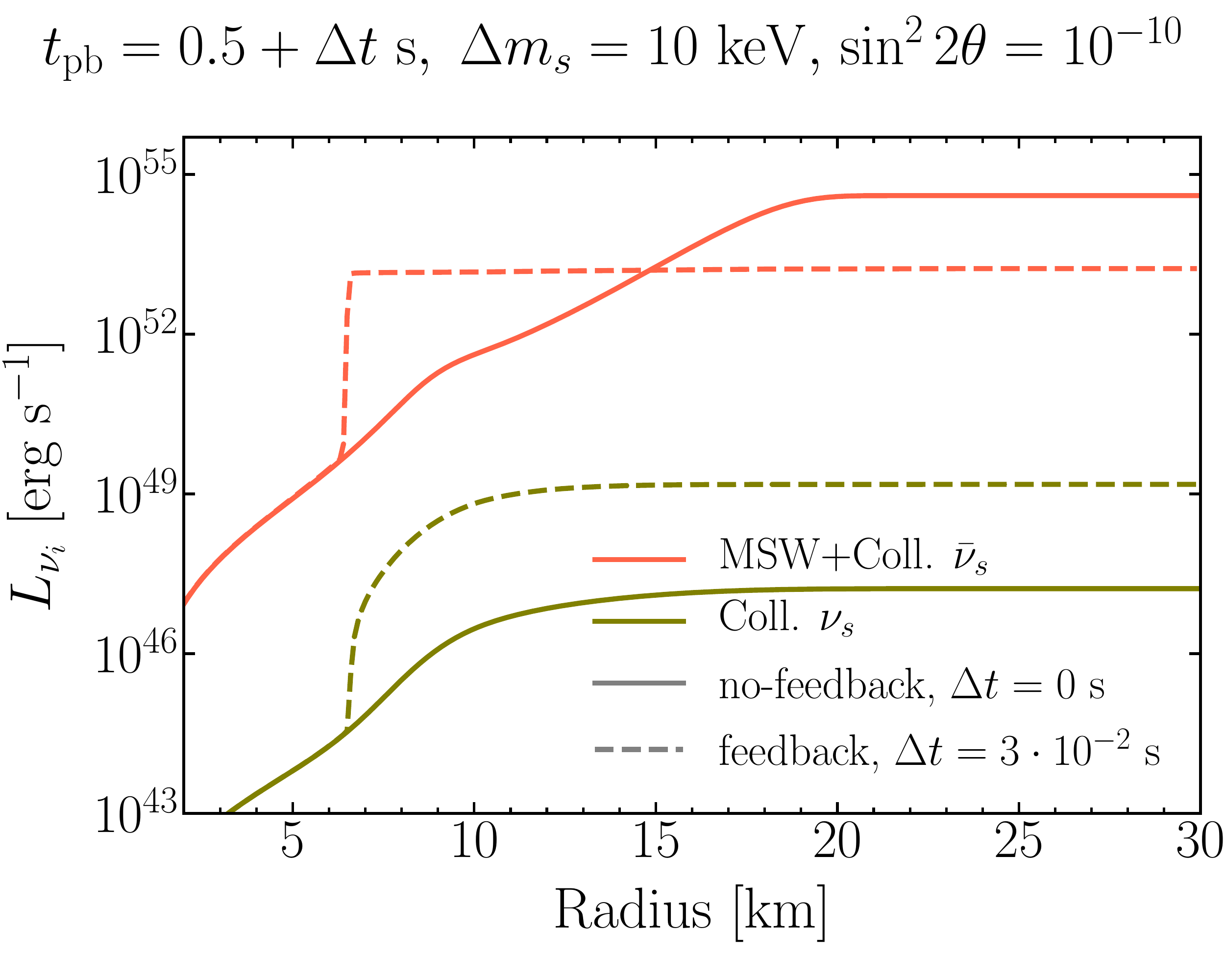}
\caption{Cumulative  luminosity of sterile neutrinos and antineutrinos as a function of the SN radius  for $t_\mathrm{pb} = 0.5$~s and mixing parameters $\Delta m_s~=~10~\mathrm{keV}$ and  $\sin^2 2\theta = 10^{-10}$ with dynamical feedback effect due to the production of sterile particles (dashed, $\Delta t = 0.03$~s) and without feedback (continuous, $\Delta t = 0$). The radial region where sterile particles are produces is affected by the feedback effects. Most of the production of sterile particles occurs within $10$--$20$~km for all the mass and mixings considered here.  The corresponding cumulative luminosity for the case without feedback is also shown in the left panel of  Fig.~\ref{fig:number_density}.}
\label{fig:lumfeed}
\end{figure}

As discussed for Figs.~\ref{fig:potential} and \ref{fig:ytau}, the temporal evolution of $V_{\mathrm{eff}}$ affects the radial range where MSW resonances occur. This is also visible in Fig.~\ref{fig:lumfeed} where the cumulative luminosities of $\nu_s$ and $\bar{\nu}_s$ have been plotted for the feedback case with $\Delta t = 0.03$~s and for the case without feedback (see also left panel of Fig.~\ref{fig:number_density}). The overall production of $\nu_s$'s is lower than the one of $\bar\nu_s$'s similarly to the case without feedback because MSW effects only take place for antineutrinos. However, given the change in shape of $V_{\mathrm{eff}}$ (see right panel of Fig.~\ref{fig:ytau}) because of the dynamical feedback, flavor conversions occur deeper inside in the SN core as time progresses.

\begin{figure}
\centering
\includegraphics[width=4.5in]{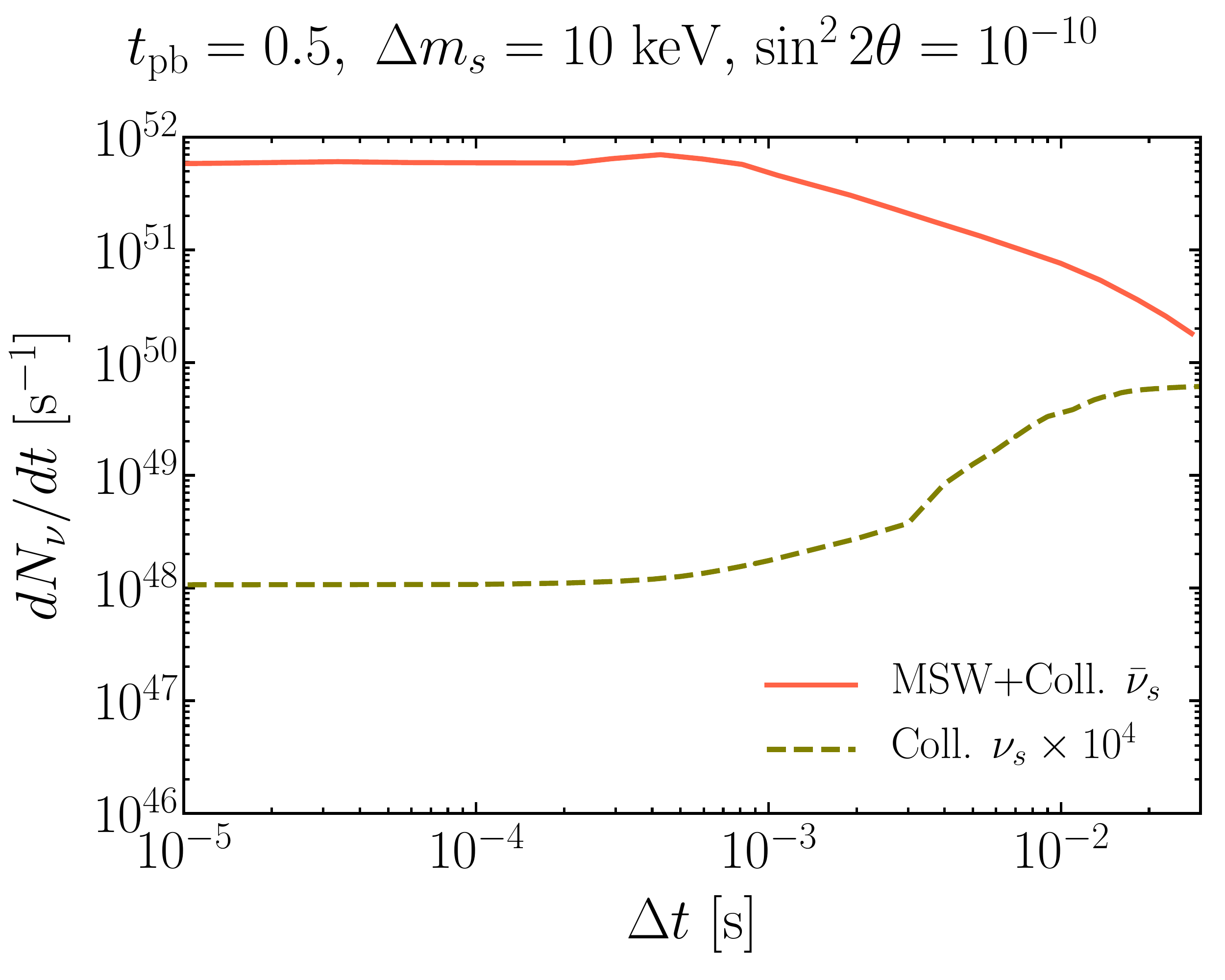}
\caption{Time evolution of the total production rate of  particles per unit time, $dN_{\nu_s}/dt$ (dashed, green line) and $dN_{\bar\nu_s}/dt$ (continuous, red line), produced up to  the neutrinosphere radius ($R_\nu = 28$~km) as a function of time for  $t_\mathrm{pb} = 0.5$~s, $(\Delta m_s, \sin^2~2\theta)=(10\ \mathrm{keV}, 10^{-10})$. As result of the dynamical feedback due to the production of sterile particles, the overall number of sterile (anti-)neutrinos is quenched after a certain $\Delta t$. }
\label{fig:fdbkrelevance}
\end{figure}
Figure~\ref{fig:fdbkrelevance} shows the temporal evolution of the production rate of sterile (anti-)neutrinos, $dN_{\nu_s}/dt$ and $dN_{\bar\nu_s}/dt$, at $R_\nu = 28$~km.  One can see that, the self-consistent implementation of the dynamical feedback effects leads to a quenching of $d(N_{\nu_s}+N_{\bar\nu_s})/dt$  after a certain time, while $d(N_{\nu_s}+N_{\bar\nu_s})/dt$  would be constant  if the dynamical feedback were not taken into account. As we will discuss later, this may have important implications in the derivation of the sterile mass-mixing bounds.

\subsection{Impact of the production of sterile particles on the lepton asymmetry}\label{sec:parspace}
To gauge the overall impact of the radial and temporal  growth of $Y_{\nu_\tau}$ under various hydrodynamical conditions, we estimate the maximum value reached by $Y_{\nu_\tau}(r)$,  $Y_{\nu_\tau}^{\rm max}$, for the three benchmark time snapshots of the $18.6\ M_\odot$ model. 
In order to do this, we  compute $dY_{\nu_\tau}/dt(\Delta t=0)$ and
the stationary $Y_{\nu_\tau}^{\rm stat}$ at different radii. 
The required time scale for $Y_{\nu_\tau}$ to reach $Y_{\nu_\tau}^{\rm stat}$ at each radius is
\begin{equation}
\Delta\tau(r)=\frac{Y_{\nu_\tau}^{\rm stat}}{\left[dY_{\nu_\tau}/dt(\Delta t=0)\right]}\ .
\end{equation}
We then compute the maximum value of $Y_{\nu_\tau}$ within $\Delta t$ as follows:
\begin{equation}
Y_{\nu_\tau}^{\rm max} = \begin{cases} Y_{\nu_\tau}^{\rm stat, max}\ &\mbox{if $\Delta \tau_{\mathrm{min}} \leq \Delta t$}\ ,\\
Y_{\nu_\tau}^{\rm stat, max} \times \frac{\Delta t}{\Delta \tau_{\mathrm{min}}}\ &\mbox{if $\Delta \tau_{\mathrm{min}} > \Delta t$}\ ,
\end{cases}
\end{equation}
where $\Delta \tau_{\mathrm{min}}$ is the minimum $\Delta t$ required by $Y_{\nu_\tau}$ to reach a stationary value, and $Y_{\nu_\tau}^{\rm stat, max}$ is the maximum of $Y_{\nu_\tau}^{\rm stat}$.

\begin{figure}
\includegraphics[width=6in]{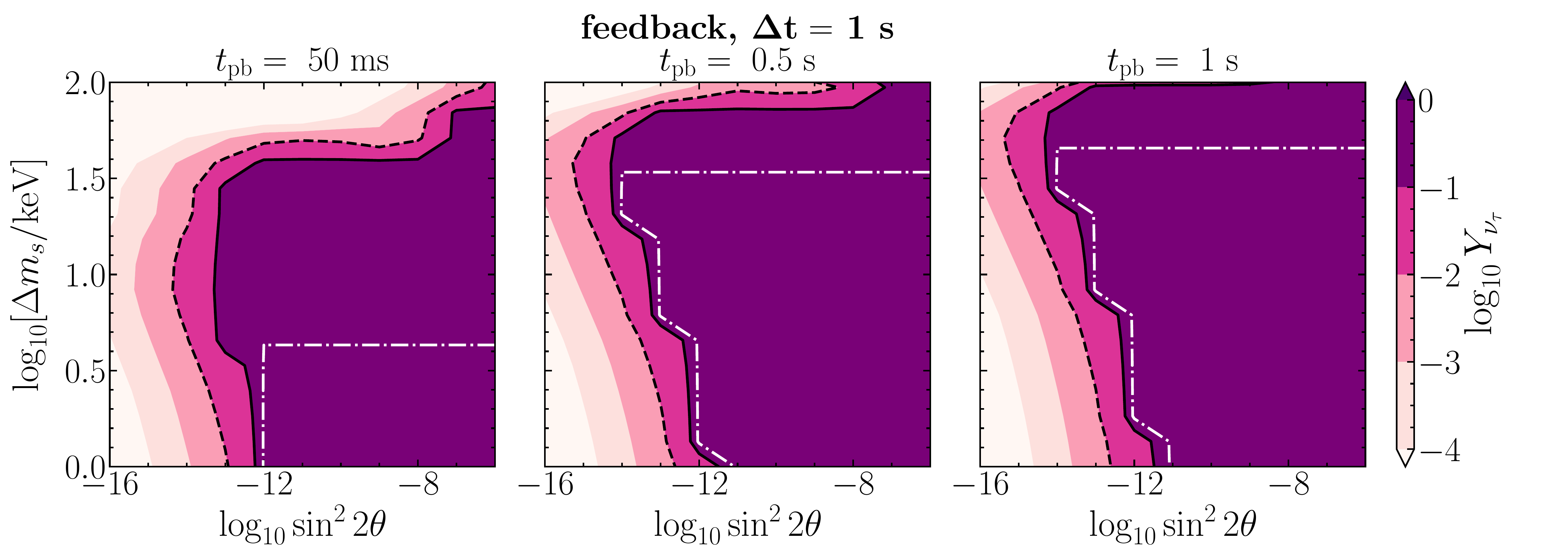}
\includegraphics[width=6in]{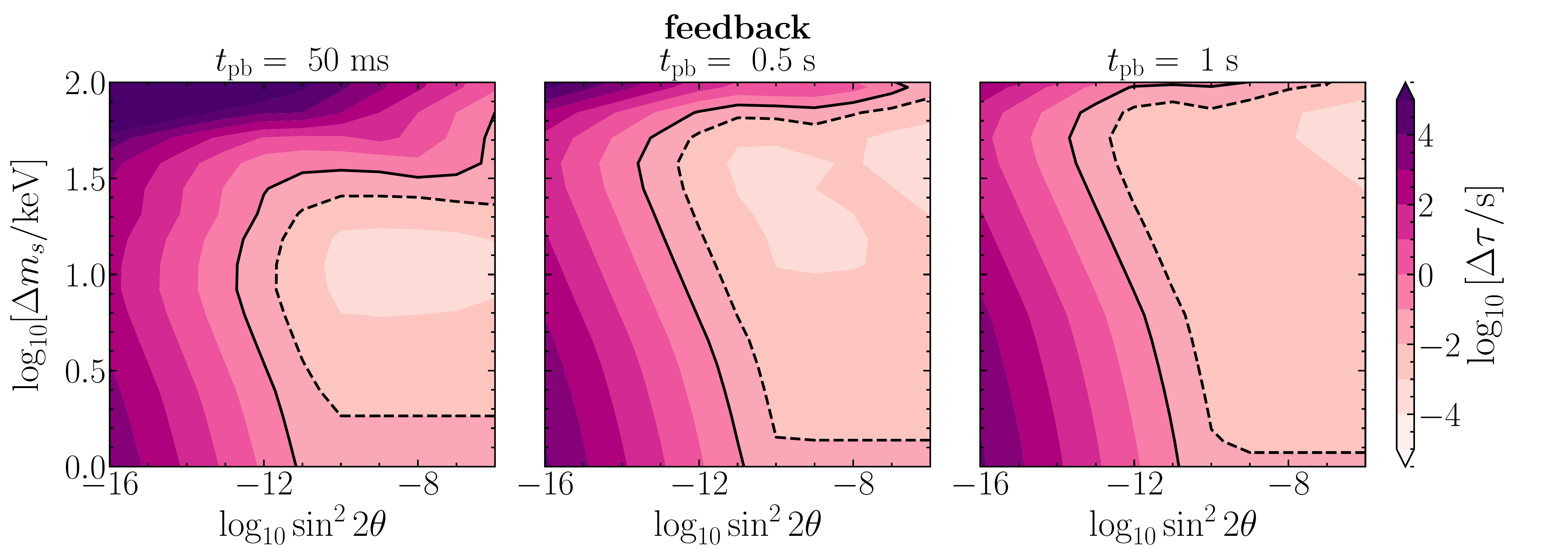}
\caption{{\it Top:} Contour plots of the  maximum of the radial profile of the stationary  $Y_{\nu_{\tau}}$ as a function of the mixing parameters $(\Delta m_s, \sin^2~2\theta)$ for  $t_\mathrm{pb} = 0.05, 0.5$ and $1$~s (on the left, middle and right, respectively) obtained by including feedback effects within $\Delta t = 1$~s. The black lines indicate $Y_{\nu_\tau} > 0.01$ (dashed) and $Y_{\nu_\tau} > 0.1$ (solid); the white dash-dotted line marks $Y_{\nu_\tau} > 0.2$. {\it Bottom:} Contour plots of the time $\Delta t$ required to reach the stationary  $Y_{\nu_{\tau}}$ (see text for details). 
}
\label{fig:2d_Y_nu_tau}
\end{figure}

Figure~\ref{fig:2d_Y_nu_tau} shows contour plots of the absolute maximum (as a function of $r$) of $Y_{\nu_{\tau}}$  in the $(\Delta m_s, \sin^2 2\theta)$ plane for  $\Delta t = 1$~s  and for $t_\mathrm{pb} = 0.05, 0.5$ and $1$~s from left to right respectively. In order to guide the eye, the continuous lines are the isocontours of  $Y_{\nu_\tau} = 0.01$ (black, dashed) and $Y_{\nu_\tau} = 0.1$ (black, solid), and $Y_{\nu_\tau} = 0.2$ (white, dot-dashed). 

We adopt $\Delta t = 1$~s, which is the typical cooling timescale for the SN core, for all three $t_\mathrm{pb}$. This is an approximation since  $\Delta t >1$~s may be needed for    $Y_{\nu_{\tau}}$ to reach its stationary configuration  depending on  $(\Delta m_s, \sin^2 2\theta, t_\mathrm{pb})$, but we see that a large range of the parameter space has $Y_{\nu_\tau}^{\rm max}=Y_{\nu_\tau}^{\rm stat}\simeq 0.2$ within $1$~s. We also show the timescale $\Delta \tau$ corresponding to $Y_{\nu_\tau}^{\rm max}$
in the bottom panel of Fig.~\ref{fig:2d_Y_nu_tau}. 
This figure clearly shows that, again, a significant portion of the parameter space
has $\Delta\tau\lesssim 1$~s. We note that although the contour shape  is in qualitative  agreement with Fig.~3 of Ref.~\cite{Raffelt:2011nc}, 
the values differ by orders of magnitude. In  agreement with Fig.~3 of Ref.~\cite{Raffelt:2011nc}, we find that very small mixing angles require large $\Delta t$  to approach the stationary solution. 
One can also see that smaller $\Delta t$ are needed to achieve the stationary $Y_{\nu_\tau}$ in a large region of the $(\Delta m_s, \sin^2 2\theta)$ parameter space as $t_\mathrm{pb}$  increases and larger values of $Y_{\nu_\tau}$ are obtained. This is due to the fact that  $Y_e$ (and therefore  $V_{\mathrm{eff}}$) is smaller in the SN core as $t_\mathrm{pb}$, allowing a wider range of energies to undergo MSW resonances. 

\newpage

\subsection{Impact of the production of sterile particles on the chemical potential}\label{sec:chempot}

Besides affecting the effective potential $V_{\rm eff}$, the generation of $Y_{\nu_{\tau}}$ due to the production of sterile neutrinos  leads to the growth of   $\nu_{\tau}$ chemical potential  $\mu_{\nu_\tau}$, which can further suppress the 
sterile neutrino production.
We compute the evolution of $\mu_{\nu_\tau}$, for fixed $t_{\mathrm{pb}}$, by finding the root of 
\begin{equation}
\label{eq:checmial_potential}
Y_{\nu_\tau} (r) - \frac{1}{n_B (r)}  \sum_{k=1}^L   \left(   \frac{d n_{\nu_\tau}}{dE_k} (r, \mu_{\nu_\tau}) -  \frac{d n_{\bar \nu_\tau}}{dE_k} (r, -\mu_{\nu_\tau})    \right) \Delta E_k = 0 \ ,
\end{equation}
with $[E_{k=1},E_{k=L}]=[E_\mathrm{min},E_\mathrm{max}]$.

\begin{figure}
\includegraphics[width=6.1in]{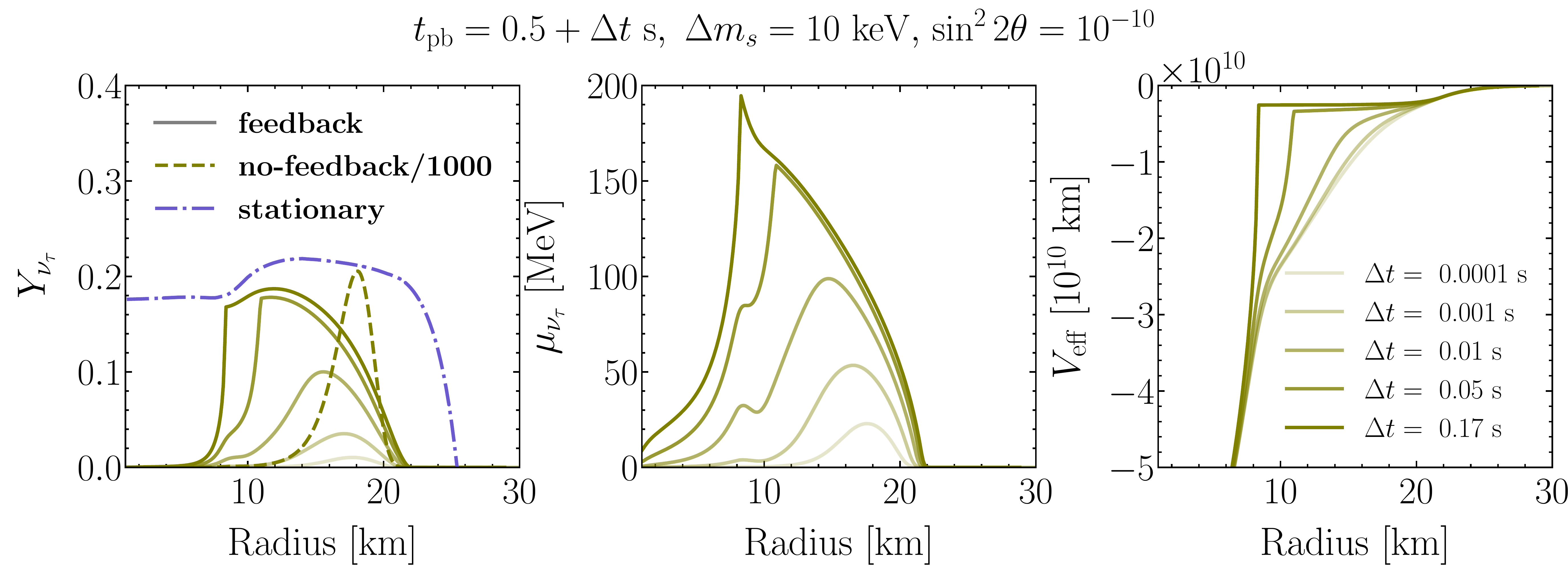}
\caption{{\it Left:} Radial profile of  $Y_{\nu_\tau}$ for the case with feedback on $Y_{\nu_\tau}$ and $\mu_{\nu_\tau}$ for $t_\mathrm{pb} = 0.5$~s, $(\Delta m_s, \sin^2~2\theta)=(10\ \mathrm{keV}, 10^{-10})$, and $\Delta t = 0.17$~s.  The dashed line represents the case without feedback ($\Delta t = 0$), the continuous lines with different hues mark the temporal evolution for the case with feedback due to the production of sterile particles.  The dot-dashed line shows $Y_{\nu_\tau}^{\rm stat}$ for $E_{\mathrm{max}}=1000$~MeV (see text for details). {\it Middle:} Radial evolution of $\mu_{\nu_\tau}$  for the case with feedback {\it Right:} Radial profile of $V_{\mathrm{eff}}$.  The dynamical feedback due to flavor conversions in sterile states leads to a considerable production of  $\mu_{\nu_\tau}$.
}
\label{fig:mu_radial}
\end{figure}
Figure \ref{fig:mu_radial} shows the temporal evolution of $Y_{\nu_\tau}$, $\mu_{\nu_\tau}$ and $V_\mathrm{eff}$ for $(\Delta m_s, \sin^2~2\theta)=(10\ \mathrm{keV}, 10^{-10})$.  
One can notice that $Y_{\nu_\tau}$ computed by including the feedback on $\mu_{\nu_\tau}$ is considerably smaller than the one calculated without any feedback.  Additionally, due to the decrease of the tau neutrino number density determined by the growth of $\mu_{\nu_\tau}$, the stationary value  $Y_{\nu_\tau}^{\rm stat}$ is not reached within $1$~s, as it was in the simulations with feedback on $V_\mathrm{eff}$ but without feedback on $\mu_{\nu_\tau}$, see Fig.~\ref{fig:ytau} for comparison.  When the feedback on $\mu_{\nu_\tau}$ is included,  the timescale necessary for reaching the final saturated value  is much longer  than in the case with feedback on $V_{\rm eff}$ only ($\Delta t\sim 0.17$~s vs.~$\Delta t\sim 0.03$~s), see Fig.~\ref{fig:ytau} for comparison. 

\begin{figure}
\includegraphics[width=6.1in]{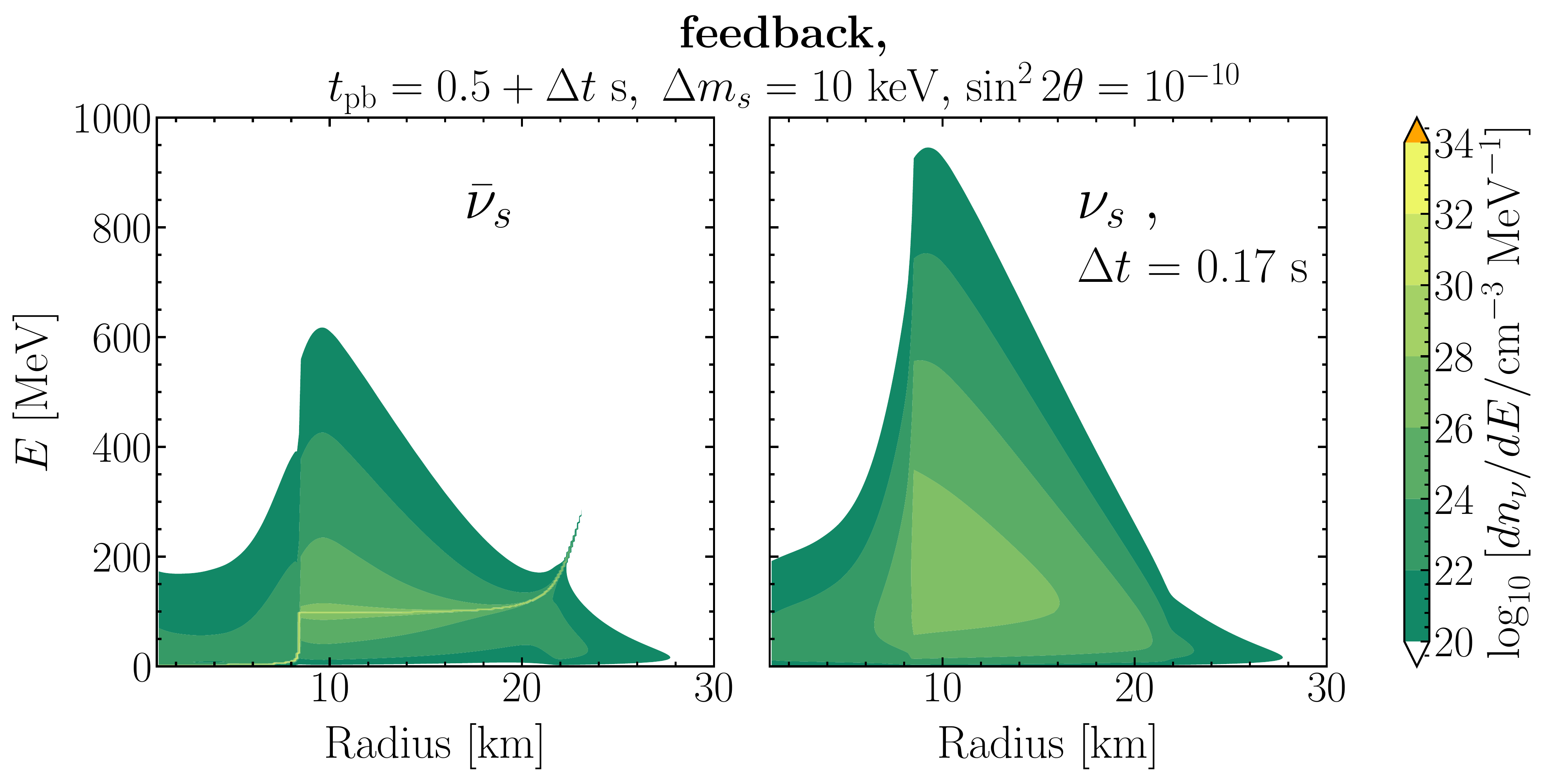}
\caption{Contour plots of the  number density of $\bar\nu_s$'s (left  panel) and $\nu_s$'s (right panel) per unit energy in the plane spanned by the distance from the SN core and the neutrino energy for $(\Delta m_s, \sin^2~2\theta)=(10\ \mathrm{keV}, 10^{-10})$, $t_\mathrm{pb} = 0.5$~s, and $\Delta t = 0.17$~s with dynamical feedback effect due to the production of sterile particles on $V_{\mathrm{eff}}$ and $\mu_{\nu_\tau}$.  See Fig.~\ref{fig:nd} for comparison with cases without the feedback ($\Delta t = 0$~s) and with feedback only on $V_{\mathrm{eff}}$.
}
\label{fig:nd_mu}
\end{figure}

Figure \ref{fig:nd_mu} shows contour plots of the sterile antineutrino (on the left) and neutrino (on the right) number densities as a function of energy and SN radius for $\Delta t = 0.17$~s, and  with feedback on $\mu_{\nu_\tau}$ and $V_\mathrm{eff}$.
By comparing Figs.~\ref{fig:nd_mu} and \ref{fig:nd}, one can notice that the resonance energy does not grow as high as in the case without the  feedback on the neutrino energy distributions. With the increase of the chemical potential, the number density of  $\bar\nu_\tau$ decreases. It results in smaller $Y_{\nu_\tau}$, as fewer antineutrinos can convert into sterile states. This leads to a smaller $V_\mathrm{eff}$ and less energetic neutrinos undergo resonances at the same radius. On the other hand, the number density of neutrinos increases due to the non-zero chemical potential.
The results of the the cumulative luminosity of sterile neutrinos and total production rate of sterile particles per unit time  are qualitatively similar to those presented in Figs.~\ref{fig:lumfeed} and \ref{fig:fdbkrelevance} when including the chemical potential feedback.

\begin{figure}
\includegraphics[width=6.1in]{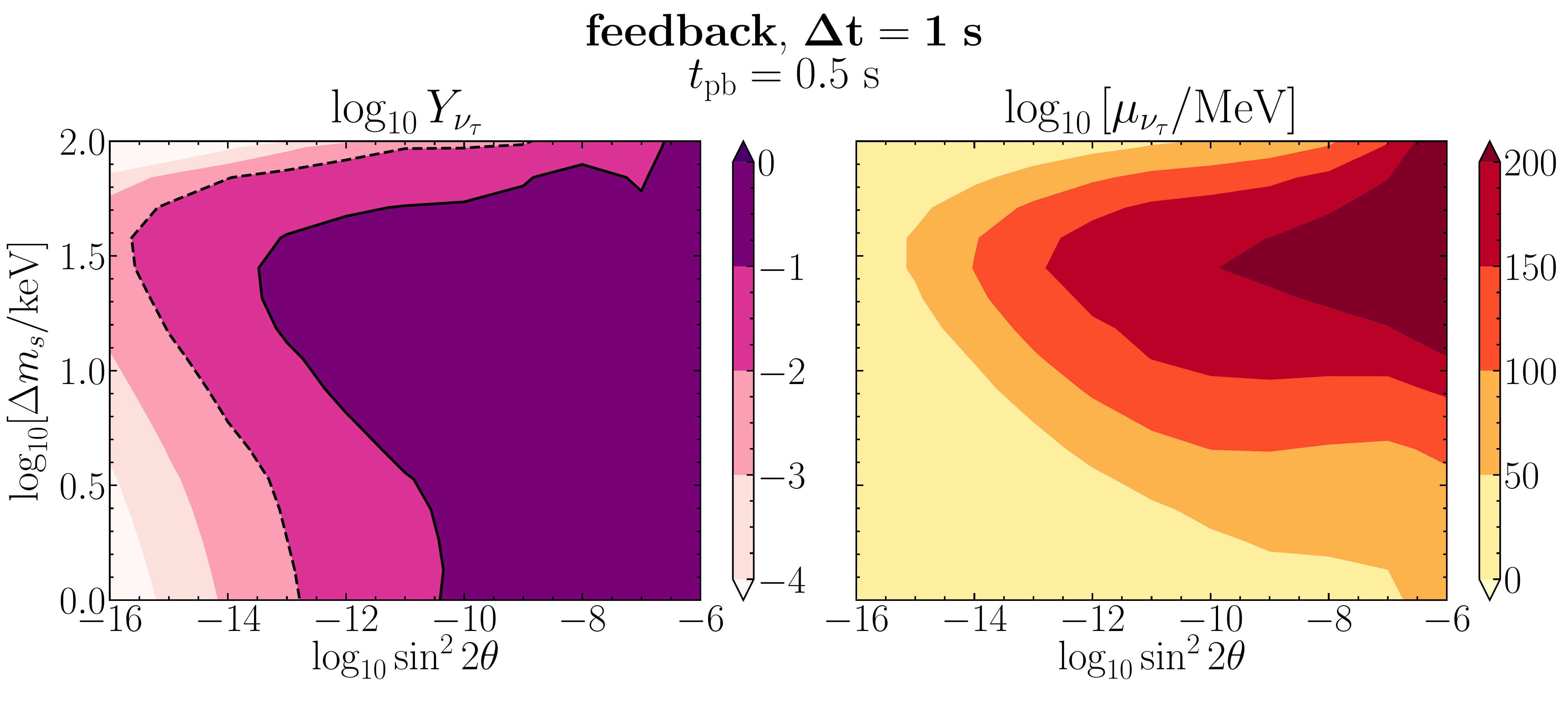}
\caption{{\it Left:} Contour plot of the maximum of the radial profile of $Y_{\nu_\tau}$ as a function of the mixing parameters ($\Delta m_s$ , $\sin^2 2\theta$) for $t_\mathrm{pb} = 0.5$~s obtained by including feedback effects on the effective matter potential and of the chemical potential on the neutrino energy distributions within $\Delta t = 1$~s. The black lines indicate $Y_{\nu_\tau}>0.01$ (dashed) and $Y_{\nu_\tau} > 0.1$ (solid). {\it Right:} Contour plot of  $\mu_{\nu_\tau}$ corresponding to the maximum of $Y_{\nu_\tau}$. The production of sterile neutrinos leads to the generation of a non-negligible $\mu_{\nu_\tau}$, which in turn has an impact on $Y_{\nu_\tau}$ and $n_{\bar \nu_\tau}$, $n_{ \nu_\tau}$. 
}
\label{fig:2D_Y_nu_tau_and_mu_nu_tau}
\end{figure}

To investigate the impact of the feedback of $\mu_{\nu_\tau} \neq 0$ on the neutrino  energy distributions and therefore on flavor conversions,  we estimate the maximum value reached by $Y_{\nu_\tau}$ and the corresponding $\mu_{\nu_\tau}$ in the $(\Delta m_s, \sin^2~2\theta)$ plane for $t_\mathrm{pb} = 0.5$~s and $\Delta t = 1$~s. By comparing the contour plot of $Y_{\nu_\tau}$ in Fig.~\ref{fig:2D_Y_nu_tau_and_mu_nu_tau} with the one in the top middle panel of  Fig.~\ref{fig:2d_Y_nu_tau}, one can see that the overall trend remains unchanged, but the maximum $Y_{\nu_\tau}$ can vary within $10 \%$. In particular, the region with  $Y_{\nu_\tau} > 0.1$ shrunk;  in that same region, a non-negligible $\mu_{\nu_\tau}$ is also produced. 

\subsection{Discussion}
The alteration of the neutrino lepton number due to the production of keV-mass sterile neutrinos was first discussed in Refs.~\cite{Hidaka:2006sg,Hidaka:2007se,Raffelt:2011nc} in the context of a one-zone model  (although Ref.~\cite{Hidaka:2006sg} investigated the $e$--$s$ mixing instead of the $\tau$--$s$ one). Also Ref.~\cite{Arguelles:2016uwb} studies the role of $\tau$--s mixing, but without considering the feedback effects on the hydrodynamical quantities. 

Reference~\cite{Raffelt:2011nc} adopted a one-zone  model with homogeneous and isotropic core, and  with  constant  temperature and baryon density.  Their Fig.~3  estimates the time required by the system to converge towards a stationary $Y_{\nu_\tau}$; despite our modeling is more sophisticated, the bottom panel of our Fig.~\ref{fig:2d_Y_nu_tau} is in qualitative agreement with  Fig.~3  of \cite{Raffelt:2011nc}; we find that $Y_{\nu_\tau}^{\rm stat}$ is reached within a much smaller $\Delta t$. 

Reference~\cite{Warren:2016slz} attempts to couple the active-sterile neutrino flavor conversion physics to the hydrodynamical  simulation by  assuming  adiabatic flavor conversions.  In Ref.~\cite{Warren:2016slz}, it is concluded that the mixing with $\nu_\tau$'s does not significantly alter the explosion physics and has no observable effect on the neutrino luminosities at early times. This is at odds with our results that include a detailed modeling of the neutrino microphysics, see bottom panel of Fig.~\ref{fig:2d_Y_nu_tau}. Most importantly, as shown in Figs.~\ref{fig:energy_spectra}, \ref{fig:number_density} and \ref{fig:lumfeed}, the adiabaticity of the flavor conversions strongly depends  on the  mixing parameters, and the collisionally enhanced production of sterile particles is non-negligible.

As highlighted in Figs.~\ref{fig:fdbkrelevance} and \ref{fig:mu_radial} (see also the left panel of Fig.~\ref{fig:ytau}), tracking the evolution of $Y_{\nu_\tau}$ and $\mu_{\nu_\tau}$ in time and radius self-consistently is important in order to derive robust constraints on the allowed $(\Delta m_s, \sin^2 2\theta)$ parameter space. Although we only investigate the sterile neutrino production on a static SN background, our results hint towards more robust bounds in the mass-mixing  parameter space of sterile neutrinos with respect to the ones  currently reported in the literature, if a self-consistent modeling of the sterile neutrino physics is included.

In order to give an idea of how existing bounds on sterile neutrino mass-mixing parameters would be modified, if a self-consistent estimation of the feedback effects were implemented,  we estimate the total energy emitted in sterile neutrinos and antineutrinos:
\begin{equation}
E_{\nu_s}(\Delta t) = \sum_{\nu, \bar{\nu}}\sum_{y=1}^N  \sum_{j=1}^{M}   \sum_{k=1}^{L} \Delta t_y \Delta r^\prime_j  \Delta E_k\  E_k \ 4 \pi r^{\prime 2}_j \ P_{\tau s}(E_{\mathrm{res}},r^\prime_j, t_y) \ \frac{dn_{\nu_\tau, \bar{\nu}_\tau}(r^\prime_j, t_y)}{dE_k} \  \frac{1}{\Delta r^\prime_j} \ ,
\end{equation}
where $[t_{y=1},t_{y=N}]=[t_{\mathrm{pb}},t_{\mathrm{pb}}+\Delta t]$, $[r^\prime_{j=1},r^\prime_{j=M}]=[1\ \mathrm{km},{R_\nu}]$, $[E_{k=1},E_{k=L}]=[E_{\mathrm{min}},E_{\mathrm{max}}]$,   $P_{\tau s}$ is  the conversion probability due to the MSW effect 
(Eq.~\ref{eq:P_MSW}) or collisions (Eq.~\ref{eq:Pxs_collisions}), and $dn_{\nu_\tau, \bar{\nu}_\tau}/dE$ has been estimated by including $\mu_{\nu_\tau}$ due to the generation of sterile particles.  The time interval $\Delta t =1$~s has been considered as representative of the  characteristic timescale of  the SN neutrino emission during the proto-neutron star cooling. We then compare $E_{\nu_s}(\Delta t)$ to the typical binding energy of a neutron star, $ E_B = 3 \times 10^{53}$~ergs.
 Note that this comparison only provides a conservative estimate of the excluded mass-mixing parameter space of sterile neutrinos. A robust bound can only be placed through a self-consistent assessment, including the dynamical feedback effects of flavor conversions in time-dependent hydrodynamical SN simulations.

\begin{figure}
\includegraphics[width=6.1in]{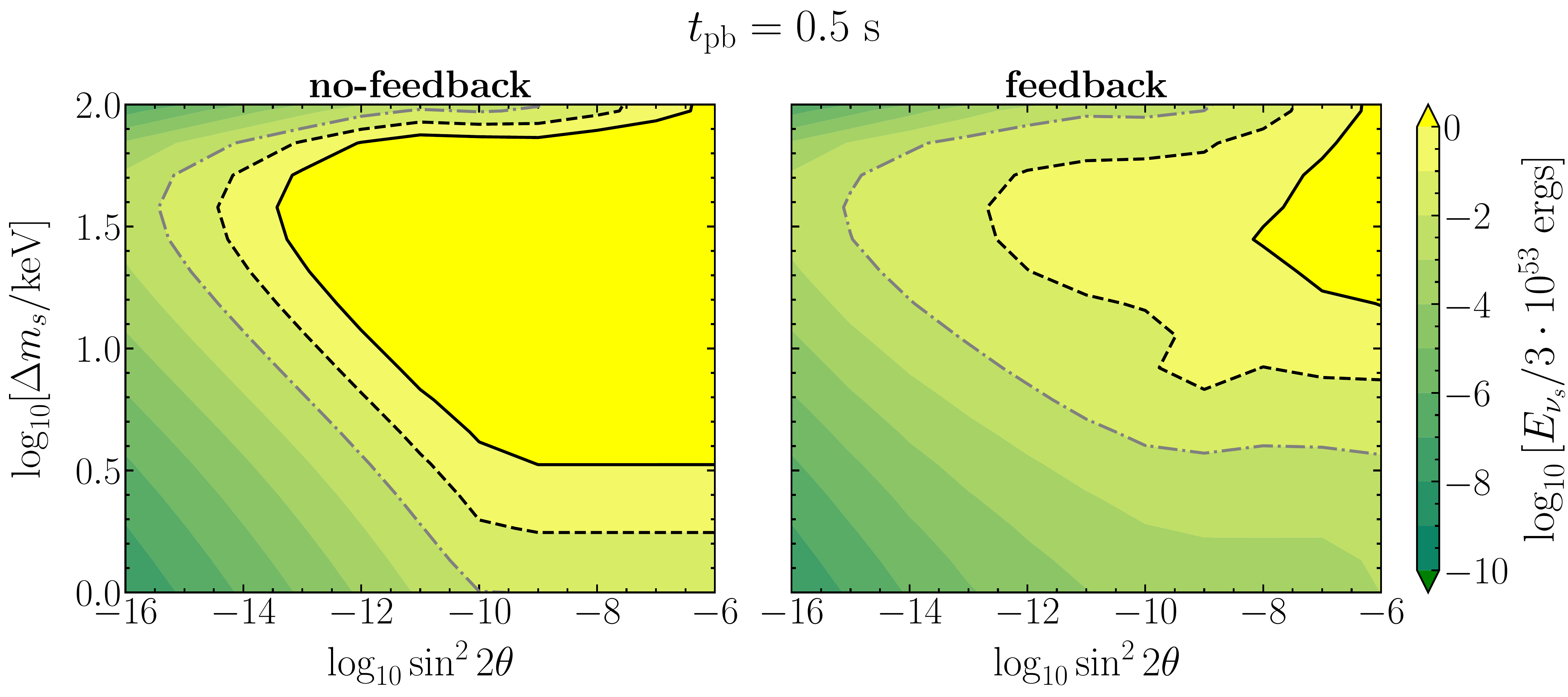}
\caption{Ratio of energy emitted in sterile (anti-)neutrinos over $\Delta t=1$~s  for $t_{\mathrm{pb}}=0.5$~s without  feedback (left panel) and with the feedback (right panel) to the benchmark gravitational energy ($E_{\nu_s}/E_B$ with $E_B = 3 \times 10^{53}~\mathrm{ergs}$).  The grey, dash-dotted line indicates  $E_{\nu_s}/E_B > 0.01$, the black dashed line stands for $E_{\nu_s}/E_B > 0.1$; the black solid line marks the excluded region, where the energy ratio $E_{\nu_s}/E_B > 1$. Although robust bounds on $(\sin^2 2\theta, \Delta m_s)$ can be derived  through a self-consistent and time-dependent estimation of the production of sterile particles only, the dynamical feedback due to flavor conversions in sterile states considerably relaxes the excluded region of the  parameter space of sterile neutrinos.} 
\label{fig:SN_bounds}
\end{figure}

Figure~\ref{fig:SN_bounds} shows the contours of  $E_{\nu_s}/E_B$  for $\Delta t =1$~s, $t_{\mathrm{pb}}=0.5$~s for the case without feedback  (left panel) and with feedback on $\mu_{\nu_\tau}$ and $Y_{\nu_\tau}$ due to the production of sterile neutrinos (right panel)\footnote{For very large mixing angles ($\sin^2 2 \theta > 10^{-10}$),  the computational time needed to estimate the feedback effects due to the production of sterile particles increases dramatically. Hence, the final values of $\mu_{\nu_\tau}$, $Y_{\nu_\tau}$, $E_{\nu_s}(\Delta t)$ for those mixing angles have been extrapolated  by a power-law. By comparison with smaller mixing angles where the runs are done for $\Delta t =1$~s, we estimate that our approximation leads to an error of a few $\%$  at most.}. 
One can easily notice that the inclusion of the dynamical feedback due to the production of sterile particles significantly affects the ``excluded''  region ($E_{\nu_s}/E_B>1$). In particular, the feedback effects seem to lead to larger region of the parameter space  to be still compatible with SNe.

\section{Outlook}\label{sec:conclusions}

Sterile neutrinos with keV mass  have the potential to dramatically affect the supernova physics generating a large neutrino lepton asymmetry. This work constitutes a step forward towards a self-consistent, radial and time-dependent modeling of the keV-mass sterile neutrino production and propagation in the supernova core. 

We develop the first self-consistent treatment of active--sterile flavor conversions, by employing stationary, but radially evolving profiles for the hydrodynamical quantities of interest. We model  the neutrino flavor evolution in the presence of  collisions in the SN core by assuming spherical symmetry and explore the development of the  $\nu_\tau$--$\bar\nu_\tau$ lepton asymmetry in radius and time as a function of ($\Delta m_s, \sin^2 2\theta$). Given the complications induced by the modeling of active--sterile conversions at nuclear densities, we  work in a two flavor framework and only  consider the mixing in the $(\nu_\tau, \nu_s)$ sector.

The  production of sterile particles with $\mathcal{O}(1-100)$~keV mass, through the $\bar\nu_\tau$--$\bar\nu_s$ MSW resonant conversions and resonantly enhanced collisions, leads to the development of a $\nu_\tau$--$\bar\nu_\tau$ lepton asymmetry. The  dynamical feedback induced on the effective matter background experienced by neutrinos is strongly affected by the production of sterile particles and is responsible for the growth in time of the $\nu_\tau$--$\bar\nu_\tau$  asymmetry, which in turn affects the (anti)neutrino energy distributions through a related non-zero chemical potential.

A scan of the mass and mixing parameter space for sterile neutrinos highlights that the potential feedback  of the sterile neutrino mixing on the $\nu_\tau$ lepton asymmetry can be larger than 0.15. This is  largely overestimated when the feedback effects are not included and would drastically affect the supernova bounds on the mass-mixing of sterile neutrinos.
 
The generation of a large $\nu_\tau$ asymmetry in the supernova core may have relevant implications for the post-bounce evolution of the proto-neutron star. For example, the average temperature of the active neutrinos in the accretion and early cooling phase may be higher due to the faster cooling of the proto-neutron star and, in turn, this may affect the neutrino heating  within a self-consistent time-dependent hydrodynamical framework. Likewise, the nucleosynthesis conditions in the neutrino-driven wind may be affected. Note that similar conclusions would also hold  in the case of $\nu_\mu$--$\nu_s$ mixing, and this may also affect the explosion mechanism subtantially~\cite{Bollig:2017lki}. 

Another observational consequence connected to the generation of sterile neutrinos with keV-mass is that the energy distributions of the active neutrinos detected on Earth will be modified. In particular, we neglect the mixing of active neutrinos of different flavors among themselves, but the production of sterile neutrinos from the $\tau$ channel will of course indirectly affect the  the other active sectors.

In conclusions, sterile neutrinos with keV mass can have a major impact on the supernova physics. Our work highlights the important implications of a self-consistent modeling of the related microphysics in order to explore the consequences of non-standard physics occurring in the core of a massive star. Only through a self-consistent modeling of the production of sterile particles, one would be able to extrapolate robust constraints on the mass and mixing parameters of these sterile and weakly-interacting particles.

\acknowledgments
We are grateful to Chris Pethick and Georg Raffelt for insightful discussions. We are also grateful to Robert Bollig and Thomas Janka for granting access to the data of the supernova model adopted in this work.
IT acknowledges support from the Villum Foundation (Project No.~13164),  the Knud H\o jgaard Foundation, and the Deutsche Forschungsgemeinschaft through Sonderforschungbereich SFB 1258 ``Neutrinos and Dark Matter in Astro- and Particle Physics'' (NDM). MRW acknowledges support from the Ministry
of Science and Technology, Taiwan under Grant No.~107-2119-M-001-038 and No.~108-2112-M-001 -010. 

\appendix
\section{Neutrino interaction rates}\label{sec:appA}
Neutrinos interact within the dense SN core. For a given interaction, not all the phase space of the outgoing particles may be available due to Pauli blocking effects. In this Appendix, in order to single out the main neutrino interaction rates  contributing to the  total rate $\Gamma_\nu$ (see Eq.~\ref{eq:nurate}), we estimate the reaction rates involving $\nu_\tau$'s and $\bar{\nu}_\tau$'s for our three selected  post-bounce times ($t_{\mathrm{pb}} = 0.05$, $0.5$ and $1$~s) in the absence of flavor conversions. We will also provide an assessment of the Pauli blocking effects of nucleons. 
An average over the neutrino energy distribution will be considered in order to have an overall picture of the most relevant processes. 

The main reactions occurring in the SN core should involve the scattering of neutrinos  on nucleons (i.e., $p + \nu_\tau \rightarrow p + \nu_\tau$ and $n + \nu_\tau \rightarrow n + \nu_\tau$):
\begin{equation}
\Gamma(p + \nu_\tau \rightarrow p + \nu_\tau) = Y_{e}\ n_B\ \frac{\int dE \sigma_{\nu p}(E) dn_{\nu_\tau}/dE}{n_{\nu_\tau}}\ , 
\end{equation}
\begin{equation}
\Gamma(n + \nu_\tau \rightarrow n + \nu_\tau) =  (1-Y_{e})\ n_B\ \frac{\int dE \sigma_{\nu n}(E) dn_{\nu_\tau}/dE}{n_{\nu_\tau}}\ ,   
\end{equation}
 where the cross sections, $\sigma_{\nu \mathrm{n}}$ and $\sigma_{\nu \mathrm{p}}$, are defined as from Ref.~\cite{Leitner:2006sp},  $dn_{\nu}/dE$ is the neutrino energy distribution introduced in Sec.~\ref{sec:nu_signal}, $Y_e$ is the electron fraction, $n_B$ the local baryon density, and $T$ the medium temperature.

The other interaction rates are derived following  Ref.~\cite{Hannestad:1997gc}, which estimated the contributions to $\Gamma_\nu$ by thermally averaging the energy-dependent interaction rate on the neutrino Fermi-Dirac distribution, closely. The neutral current (NC) scattering of neutrinos among themselves is governed by the following rates
\begin{equation}
\Gamma(\nu_\tau  \bar{\nu}_{\tau} \rightarrow \nu_{e} \bar{\nu}_{e}) = \frac { 4 } { \pi ^ { 3 } } G _ { F } ^ { 2 } T ^ { 5 }\ ,
\end{equation}
and 
\begin{equation}
\Gamma(\nu_{\tau} \nu_{e} \rightarrow \nu_{\tau} \nu_{e},\ \nu_{\tau} \bar{\nu}_{e} \rightarrow \nu_{\tau} \bar{\nu}_{e}) =  4  \Gamma(\nu_x  \bar{\nu}_{\tau} \rightarrow \nu_{e} \bar{\nu}_{e}) = \frac { 16 } { \pi ^ { 3 } } G _ { F } ^ { 2 } T ^ { 5 }\ .
\end{equation}

The $\nu$--$\nu$ scattering leading to $e^+$--$e^-$ pair production has instead an interaction rate defined as
\begin{equation}
\Gamma(\nu_{\tau} \bar{\nu}_{\tau} \rightarrow e^{+} e^{-}) = \Gamma(\nu_\tau  \bar{\nu}_{\tau} \rightarrow \nu_{e} \bar{\nu}_{e}) \left( C _ { V , e } ^ { 2 } + C _ { A , e } ^ { 2 } \right) \eta _ { e } ^ { 4 } e ^ { - \eta _ { e } } / 12\ ,
\end{equation}
where $\eta_e = \mu_e/T$ is the degeneracy parameter and $C_{V,e} (C_{A,e}) = 1/2 (- 1/2 + 2 \sin^{2} \theta_W)$ is the vector (axial) coupling constant with $\sin^{2} \theta_{\mathrm { W }} = 0.23$. 
The related interaction rate for $\nu_{\tau} e^{-} \rightarrow \nu_{\tau} e^{-}$ is 
\begin{equation}
\Gamma(\nu_{\tau} e^{-} \rightarrow \nu_{\tau} e^{-})  = \Gamma(\nu_\tau  \bar{\nu}_{\tau} \rightarrow \nu_{e} \bar{\nu}_{e}) \left( C _ { V , e } ^ { 2 } + C _ { A , e } ^ { 2 } \right) 3 \eta _ { e } ^ { 2 }\ .
\end{equation}

The thermally averaged rate for the Bremsstrahlung processes $\nu_{\tau} \bar{\nu}_{\tau} N N \rightarrow N N$ is: 
\begin{equation}
\Gamma(\nu_{\tau} \bar{\nu}_{\tau} n n \rightarrow n n)  = \frac { 3 C _ { A } ^ { 2 } G _ { F } ^ { 2 } } { \pi } \left( 1-Y_e\right)n _ { B } T ^ { 2 } \frac { \Gamma _ { \sigma } / T } { 20 \pi } \xi\ ,
\end{equation}
and 
\begin{equation}
\Gamma(\nu_{\tau} \bar {\nu}_{\tau} p p \rightarrow p p) = \frac { 3 C _ { A } ^ { 2 } G _ { F } ^ { 2 } } { \pi } Y_e n_{ B } T ^ { 2 } \frac { \Gamma _ { \sigma } / T } { 20 \pi } \xi\ .
\end{equation}
In the two Bremsstrahlung  rates introduced above,  $\xi$ is the dimensionless mean free path and it has been assumed to be $1$ for simplicity~\cite{Hannestad:1997gc}. 
The spin fluctuation rate ($\Gamma_{\sigma}$) is given by
\begin{equation}
\Gamma _ { \sigma } = \frac { 8 \sqrt { 2 \pi } \alpha _ { \pi } ^ { 2 } } { 3 \pi ^ { 2 } } \eta _ { * } ^ { 3 / 2 } \frac { T ^ { 2 } } { m _ { N } },
\end{equation}
where  $\eta _ { * } = p_F^2/(2m_N T)$ is the degeneracy parameter for the nucleons, $p_F= \sqrt{2 m_N \mu_N}$ the Fermi momentum of nucleons with $m_N$  their mass, and  $\alpha _ { \pi } = \left( f 2 m_{N}/m_{\pi} \right)^{2}/4 \pi \approx 15$  with  $f \approx 1$.

\begin{figure}
\centering
\includegraphics[scale=0.4]{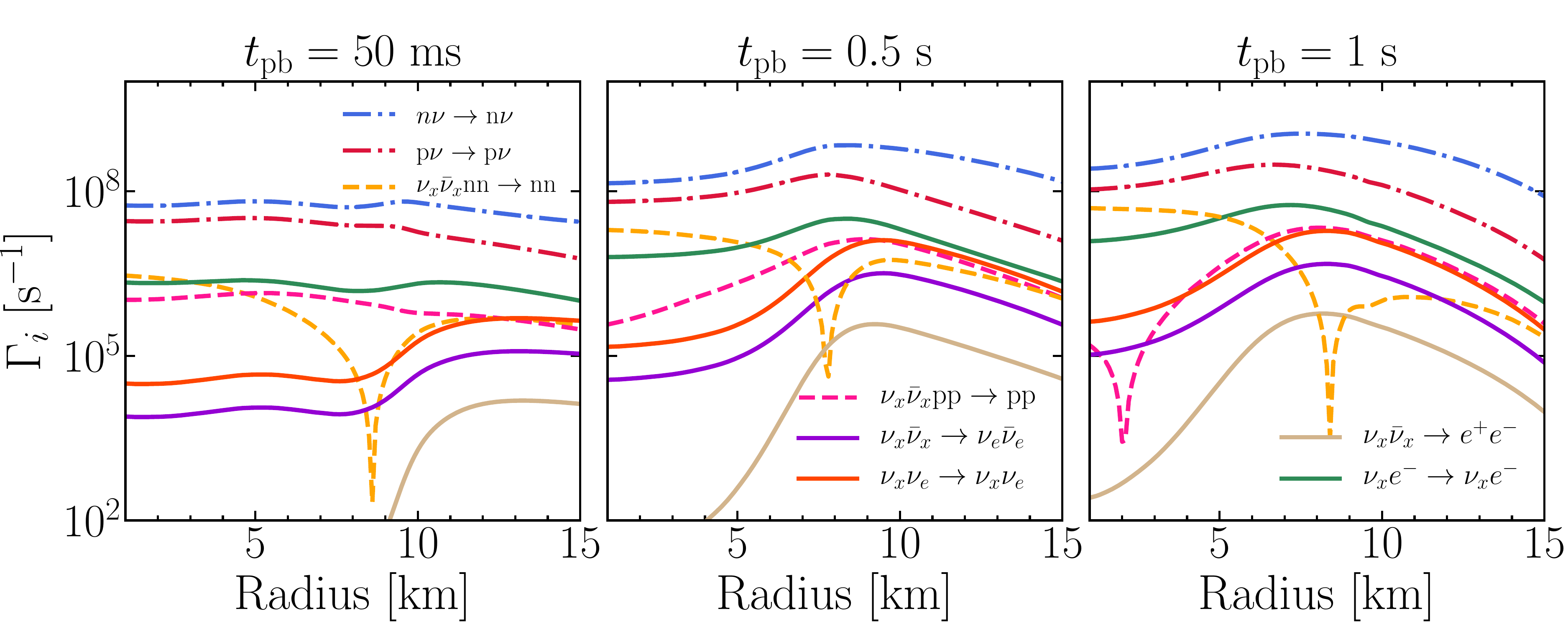}
\caption{Neutrino interaction rates averaged on the neutrino energy distribution as a function of the distance from the SN core for $t_{\mathrm{pb}} = 0.05, 0.5$ and $1$~s (from left to right, respectively). The dominant interaction channels are the NC $\nu$--$N$ scatterings for all post-bounce times. The dips present below $\sim$~10~km, in the $ \nu_{\tau} \bar\nu_{\tau} n n \rightarrow n n$ lines come from $\mu_\mathrm{n}$ crossing zero.}
\label{fig:NC_n}
\end{figure}

Figure~\ref{fig:NC_n} shows the neutrino interaction rates introduced above averaged over the neutrino energy distribution as a function of the distance from the SN core. The different panels correspond to $t_{\mathrm{pb}} = 0.05, 0.5$ and $1$~s from left to right, respectively. The various reaction rates have been plotted with different colors as indicated in the legend. The NC scatterings of neutrinos on nucleons, $p + \nu_\tau \rightarrow p + \nu_\tau$ and $n + \nu_\tau \rightarrow n + \nu_\tau$, provide the main contribution to the total neutrino reaction rate $\Gamma_\nu$ followed by the $\nu$ scattering on electrons. Notably since the $\nu$--$e$ scattering should mainly contribute in the proximity of the neutrino decoupling radius (i.e., when neutrinos are free-streaming) and all others rates are sub-leading, we only include  the energy-dependent reaction rates $N + \nu_\tau \rightarrow N + \nu_\tau$ in the estimation of the  the sterile neutrino collisional production. Moreover, we assume that the scattering on nucleons is elastic up to  $E_{\mathrm{max}}=1000$~MeV.  In fact, as discussed in Sec.~\ref{sec:Y_tau}, the  contribution to the sterile production coming from collisions for neutrino energies  above $400$--$500$~MeV is only relevant  for very high masses of sterile neutrinos (above $100$~keV).

In the dense SN core, for a given reaction, not all final states are allowed because of Pauli blocking effects~\cite{Raffelt:1996wa,Lamb:1976ac}. 
In fact, the Pauli blocking forbids the scattering of fermions into occupied states, leading to a suppression of the neutrino  interaction rate. In the following, for the sake of simplicity, we assume that the distribution function of nucleons does not change as a result of the scattering of nucleons on neutrinos (i.e.,~$E_{\nu} \ll m_N$). This assumption leads to a simplified estimation of the  Pauli blocking  factor of nucleons that suits well our purposes. In the derivation of the Pauli blocking effects, we closely follow Refs.~\cite{Raffelt:1996wa,Bruenn:1985en} and focus on the main reaction $N + \nu_\tau \rightarrow N + \nu_\tau$:  
\begin{equation}
\label{eq:Pauli_blocking_factor}
F_{\nu, N}(E) = \left(1 - f(E)\right) \int_{0}^{\infty} \frac{2}{2 \pi^2 n_{N}} {dp_N\ p_N^2}\ f(E_N)\ \left[1 - f(E_N)\right]\ ,
\end{equation} 
where $E_N = \sqrt{p_N^2 + m^2_N} + U_\star$ is the nucleon energy, with $p_N$ the nucleon momentum, $m_N$ the effective nucleon mass, $U_\star$ the nucleon mean field potential, and $f(E_i)$ is the Fermi-Dirac energy distribution.

\begin{figure}
\centering
\includegraphics[scale=0.4]{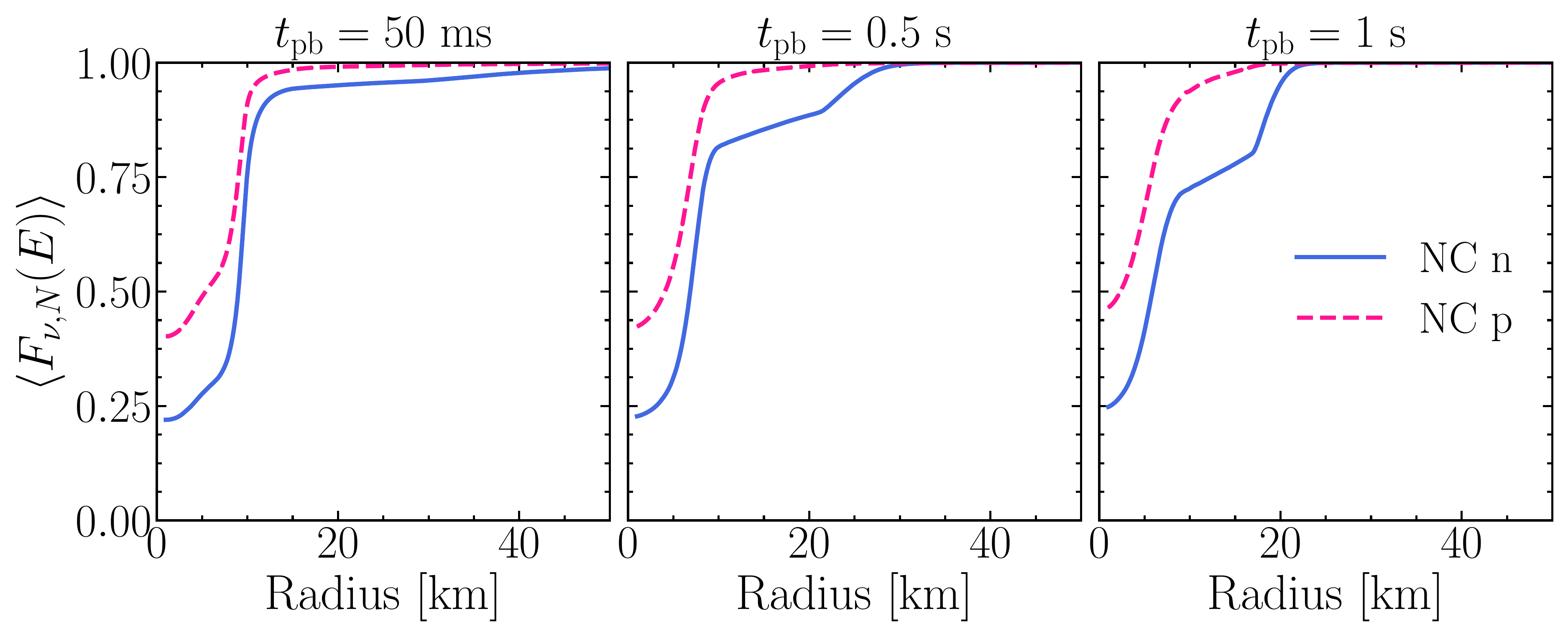}
\caption{Pauli blocking factor of nucleons averaged over the neutrino energy distribution for  the NC scattering of $\nu_\tau$ and $\bar{\nu}_\tau$  as a function of the distance from the SN core for $t_{\mathrm{pb}} = 0.05, 0.5$ and $1$~s (from left to right, respectively). The Pauli blocking effects are larger for $\nu$--$n$ interactions. The Pauli blocking becomes negligible in the proximity of the neutrino decoupling region.}
\label{fig:Pauliblock}
\end{figure}

Figure~\ref{fig:Pauliblock} shows the radial dependence of $F_{\nu, N}(E)$  averaged over the neutrino energy distribution for $t_{\mathrm{pb}} = 0.05, 0.5$ and $1$~s from left to right, respectively. Notably the Pauli blocking effect is slightly larger for the $\nu$--$n$ channel. Moreover,  $\langle F_{\nu, N}(E) \rangle \rightarrow 1$ as neutrinos approach the decoupling radius, as expected. 

The Pauli blocking factor $F_{\nu, N}(E)$  modifies the total collision rate of neutrinos and antineutrinos~\footnote{Notably, the collision rate will be modified in a different way for neutrinos and antineutrinos, if they have a non-zero chemical potential as $\nu_e$ and $\bar{\nu}_e$. In fact, the Fermi-Dirac distribution depends on the neutrino chemical potential, and $\mu_{\nu} = -\mu_{\bar\nu}$.}.
The overall interaction rate   is then defined as:
\begin{equation}
\label{eq:pauli_blocked_gamma}
\Gamma_{\nu_\tau} (E) = F_{\nu, \mathrm{p}}(E) \sigma_{\nu, \mathrm{p}}(E) Y_{\mathrm{p}} + F_{\nu, \mathrm{n}}(E) \sigma_{\nu, \mathrm{n}}(E) Y_{\mathrm{n}}\ ,
\end{equation} 
with $Y_p = Y_e$ ($Y_n= 1-Y_p$)   the proton (neutron) fraction.

\section{Neutrino conversions in the trapping regime}\label{sec:AppB}

\begin{figure}[b]
\centering
\includegraphics[scale=0.5]{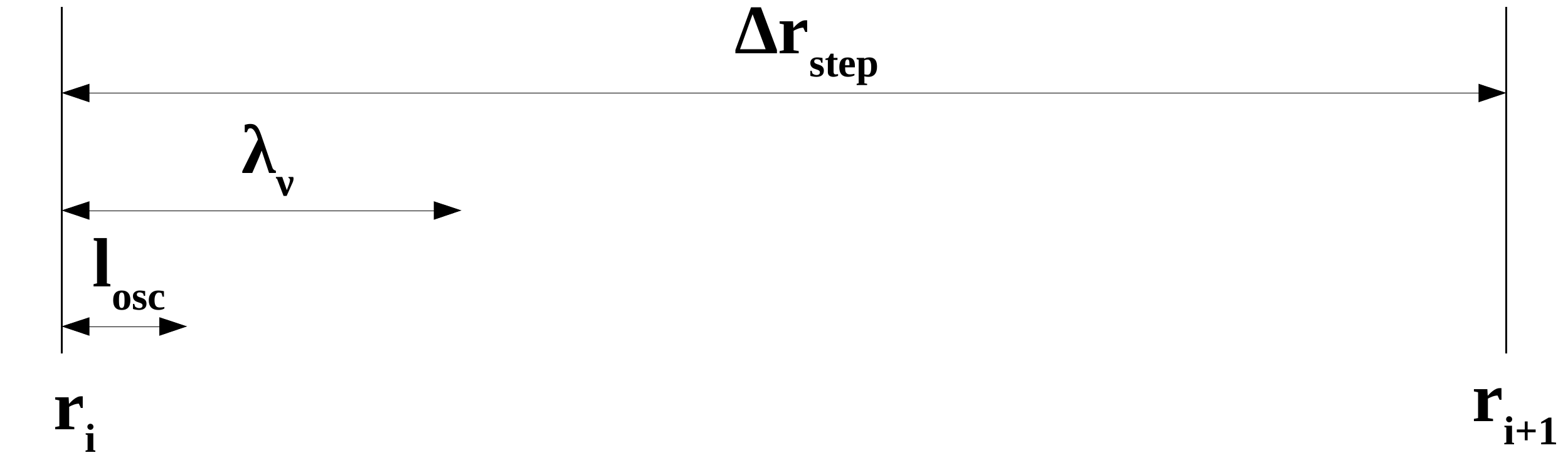}
\caption{Diagram representing a comparison among $\Delta r_{\mathrm{step}}$, $\lambda_\nu$, and $l_{\mathrm{osc}}$ within one SN shell. Multiple collisions occur within one SN shell.  Notice that the diagram is not in scale. For example, deep inside the SN core for $r \sim 5~\mathrm{km}$, $(\Delta m_s, \sin^2 2\theta) = (10~\mathrm{keV}, 10^{-6}$), and neutrino energy $E = 100~\mathrm{MeV}$, we have $l_\mathrm{osc} \sim 10^{-10}~\mathrm{km}$, $\lambda_\nu \sim 10^{-4} ~\mathrm{km}$, and $\Delta r_{\mathrm{step}}= 0.1$~km. }
\label{mutilple_shells}
\end{figure}

In the SN core, neutrinos are trapped and sterile particles can be produced because of collisions, as illustrated in Sec.~\ref{sec:collisions}. Within each SN shell of width $\Delta r_{\mathrm{step}}$, neutrinos may undergo multiple collisions if $\lambda_\nu \ll \Delta r_{\mathrm{step}}$ as illustrated in Fig.~\ref{mutilple_shells}. As a consequence, we  estimate the overall production of sterile neutrinos through each SN shell by adopting a recursive method.  

For the sake of simplicity, we assume that the number density of neutrinos is constant through the each SN shell of width $\Delta r_{\mathrm{step}}$. This is a reasonable assumption given that $\Delta r_{\mathrm{step}} \simeq \mathcal{O}(10^{-2}-10^{-1})$~km. We can then sub-divide $\Delta r_{\mathrm{step}}$ in multiple sub-shells each of width $\lambda_\nu \ll \Delta r_{\mathrm{step}}$ as shown in Fig.~\ref{mutilple_shells}.

If no sterile neutrinos are present in the first sub-shell, then the number density of sterile particles produced in the first sub-shell of width  $\lambda_\nu$ for a fixed energy $E$ is 
\begin{equation}
n_{\nu_s}^{(1)} =  n_{\nu_\tau}^{(0)}  \langle P_{\tau s} \rangle\ ,
\end{equation}
with $\langle P_{\tau s} \rangle$ defined as in Eq.~\ref{eq:Pxs_collisions} and $n_{\nu_\tau}^0$ denoting the density of $\nu_\tau$ at the beginning of the first SN sub-shell.  

Similarly, in the second sub-shell of width $\lambda_\nu$, the number density of sterile particles is
\begin{equation}
n_{s}^{(2)} =  n_{\nu_\tau}^{(1)} \langle P_{\tau s} \rangle + n_{\nu_s}^{(1)} (1 - \sin^2 2 \widetilde{\theta} \langle P_{\tau s} \rangle)\ ,
\end{equation}
where $\widetilde{\theta}$ is the mixing angle in matter assuming a constant matter potential within $\lambda_{\nu}$, and it is defined as $\sin^2 2 \widetilde{\theta} = 4 \langle P_{\tau s} (E, r)\rangle$.  The term $(1 -  \sin^2 2 \widetilde{\theta} \langle P_{\tau s} \rangle)$ accounts for a possibility of reconversion of sterile neutrinos in active ones; $n_{\nu_\tau}^{(1)} \approx n_{\nu_\tau}^{(0)}$ due to the assumption that the active neutrinos are trapped and their number density within one SN shell is constant.
At the end of the SN shell of width $\Delta r_{\mathrm{step}}$, the number density  of sterile particles is
\begin{equation}
n_{\nu_s}^{(n)} = \sum_{k=1}^{k=n}  n_{\nu_\tau}^{(0)}  \langle P_{\tau s} \rangle (1 -   \sin^2 2\widetilde{\theta} \langle P_{\tau s}\rangle )^{k - 1}\ .
\end{equation}

The overall conversion factor within a SN shell of width $\Delta r_{\mathrm{step}}$ is then 
\begin{equation}
\label{eq:Pxs_n_coll}
\mathcal{P}_{\tau s} (n) = \sum_{k=1}^{k=n}  \langle P_{\tau s} \rangle (1 -   \sin^2 2\widetilde{\theta} \langle P_{\tau s}\rangle )^{k - 1}\ .
\end{equation}

The effective conversion probability follows from Eq.~\ref{eq:Pxs_n_coll}:   
\begin{equation}
\label{eq:multiple1}
P_{\tau s} (E, n) = \frac{1}{n}\sum_{k=1}^{n}   \langle P_{\tau s} \rangle (1 -  \sin^2 2\widetilde\theta \langle P_{\tau s}\rangle )^{k - 1} = \frac{1 -  (1 - \sin^2 2\widetilde\theta \langle P_{\tau s}\rangle )^{n}}{n \langle P_{\tau s} \rangle}\ .
\end{equation}

The energy distribution of sterile (anti-)neutrinos  produced through collisions until the $\nu_\tau$ neutrinosphere radius is 
\begin{equation}
\label{eq:res11a}
\left(\frac{d\mathcal{N}}{dEdt}\right)_{s,\mathrm{coll}} = \sum_{i=1}^{N}  \Delta V_i    \frac{dn_{\nu_\tau}}{dE}(r^\prime_i)  P_{\tau s}(E,n, r^\prime_i) \Gamma_{\nu}(r^\prime_i) \ 
\end{equation}
with $\Delta V_i = 4\pi r_i^{\prime 2} \Delta r_{\mathrm{step},i}$; this is equivalent to 
\begin{equation}
\label{eq:res111a}
\left(\frac{d\mathcal{N}}{dEdt}\right)_{s,\mathrm{coll}} = \sum_{i=1}^{N}  \Delta V_i   \frac{dn_{\nu_\tau}}{dE}(r^\prime_i) n P_{\tau s}(E,n, r^\prime_i) \Delta r_{\mathrm{step},i}^{-1} \ .
\end{equation}

\bibliographystyle{JHEP}
\bibliography{sterile_nus}

\providecommand{\href}[2]{#2}\begingroup\raggedright\begin{thebibliography}{10}

\bibitem{Mirizzi:2015eza}
A.~Mirizzi, I.~Tamborra, H.-T. Janka, N.~Saviano, K.~Scholberg, R.~Bollig
  et~al., \emph{{Supernova Neutrinos: Production, Oscillations and Detection}},
  \href{http://dx.doi.org/10.1393/ncr/i2016-10120-8}{\emph{Riv. Nuovo Cim.}
  {\bf 39} (2016) 1--112}, [\href{https://arxiv.org/abs/1508.00785}{{\tt
  1508.00785}}].

\bibitem{Janka:2012wk}
H.-T. Janka, \emph{{Explosion Mechanisms of Core-Collapse Supernovae}},
  \href{http://dx.doi.org/10.1146/annurev-nucl-102711-094901}{\emph{Ann. Rev.
  Nucl. Part. Sci.} {\bf 62} (2012) 407--451},
  [\href{https://arxiv.org/abs/1206.2503}{{\tt 1206.2503}}].

\bibitem{Janka:2017vcp}
H.-T. {Janka}, \emph{{Neutrino-Driven Explosions, Handbook of Supernovae}},
  p.~1095.
\newblock 2017.
\newblock \href{https://arxiv.org/abs/[1702.08825]}{{\tt [1702.08825]}}.

\bibitem{Kotake:2012nd}
K.~Kotake, K.~Sumiyoshi, S.~Yamada, T.~Takiwaki, T.~Kuroda, Y.~Suwa et~al.,
  \emph{{Core-Collapse Supernovae as Supercomputing Science: a status report
  toward 6D simulations with exact Boltzmann neutrino transport in full general
  relativity}}, \href{http://dx.doi.org/10.1093/ptep/pts009}{\emph{PTEP} {\bf
  2012} (2012) 01A301}, [\href{https://arxiv.org/abs/1205.6284}{{\tt
  1205.6284}}].

\bibitem{Burrows:2012ew}
A.~Burrows, \emph{{Colloquium: Perspectives on core-collapse supernova
  theory}}, \href{http://dx.doi.org/10.1103/RevModPhys.85.245}{\emph{Rev. Mod.
  Phys.} {\bf 85} (2013) 245}, [\href{https://arxiv.org/abs/1210.4921}{{\tt
  1210.4921}}].

\bibitem{Chakraborty:2016yeg}
S.~Chakraborty, R.~Hansen, I.~Izaguirre and G.~G. Raffelt, \emph{{Collective
  neutrino flavor conversion: Recent developments}},
  \href{http://dx.doi.org/10.1016/j.nuclphysb.2016.02.012}{\emph{Nucl. Phys.}
  {\bf B908} (2016) 366--381}, [\href{https://arxiv.org/abs/1602.02766}{{\tt
  1602.02766}}].

\bibitem{Dasgupta:2011jf}
B.~Dasgupta, E.~P. O'Connor and C.~D. Ott, \emph{{The Role of Collective
  Neutrino Flavor Oscillations in Core-Collapse Supernova Shock Revival}},
  \href{http://dx.doi.org/10.1103/PhysRevD.85.065008}{\emph{Phys. Rev.} {\bf
  D85} (2012) 065008}, [\href{https://arxiv.org/abs/1106.1167}{{\tt
  1106.1167}}].

\bibitem{OConnor:2018sti}
E.~O'Connor et~al., \emph{{Global Comparison of Core-Collapse Supernova
  Simulations in Spherical Symmetry}},
  \href{http://dx.doi.org/10.1088/1361-6471/aadeae}{\emph{J. Phys.} {\bf G45}
  (2018) 104001}, [\href{https://arxiv.org/abs/1806.04175}{{\tt 1806.04175}}].

\bibitem{Janka:2016fox}
H.-T. Janka, T.~Melson and A.~Summa, \emph{{Physics of Core-Collapse Supernovae
  in Three Dimensions: a Sneak Preview}},
  \href{http://dx.doi.org/10.1146/annurev-nucl-102115-044747}{\emph{Ann. Rev.
  Nucl. Part. Sci.} {\bf 66} (2016) 341--375},
  [\href{https://arxiv.org/abs/1602.05576}{{\tt 1602.05576}}].

\bibitem{Sawyer:2015dsa}
R.~F. Sawyer, \emph{{Neutrino cloud instabilities just above the neutrino
  sphere of a supernova}},
  \href{http://dx.doi.org/10.1103/PhysRevLett.116.081101}{\emph{Phys. Rev.
  Lett.} {\bf 116} (2016) 081101},
  [\href{https://arxiv.org/abs/1509.03323}{{\tt 1509.03323}}].

\bibitem{Sawyer:2008zs}
R.~F. Sawyer, \emph{{The multi-angle instability in dense neutrino systems}},
  \href{http://dx.doi.org/10.1103/PhysRevD.79.105003}{\emph{Phys. Rev.} {\bf
  D79} (2009) 105003}, [\href{https://arxiv.org/abs/0803.4319}{{\tt
  0803.4319}}].

\bibitem{Izaguirre:2016gsx}
I.~Izaguirre, G.~G. Raffelt and I.~Tamborra, \emph{{Fast Pairwise Conversion of
  Supernova Neutrinos: A Dispersion-Relation Approach}},
  \href{http://dx.doi.org/10.1103/PhysRevLett.118.021101}{\emph{Phys. Rev.
  Lett.} {\bf 118} (2017) 021101},
  [\href{https://arxiv.org/abs/1610.01612}{{\tt 1610.01612}}].

\bibitem{Abbar:2018beu}
S.~Abbar and M.~C. Volpe, \emph{{On Fast Neutrino Flavor Conversion Modes in
  the Nonlinear Regime}},
  \href{http://dx.doi.org/10.1016/j.physletb.2019.02.002}{\emph{Phys. Lett.}
  {\bf B790} (2019) 545--550}, [\href{https://arxiv.org/abs/1811.04215}{{\tt
  1811.04215}}].

\bibitem{Dasgupta:2016dbv}
B.~Dasgupta, A.~Mirizzi and M.~Sen, \emph{{Fast neutrino flavor conversions
  near the supernova core with realistic flavor-dependent angular
  distributions}},
  \href{http://dx.doi.org/10.1088/1475-7516/2017/02/019}{\emph{JCAP} {\bf 1702}
  (2017) 019}, [\href{https://arxiv.org/abs/1609.00528}{{\tt 1609.00528}}].

\bibitem{Capozzi:2018clo}
F.~Capozzi, B.~Dasgupta, A.~Mirizzi, M.~Sen and G.~Sigl, \emph{{Collisional
  triggering of fast flavor conversions of supernova neutrinos}},
  \href{http://dx.doi.org/10.1103/PhysRevLett.122.091101}{\emph{Phys. Rev.
  Lett.} {\bf 122} (2019) 091101},
  [\href{https://arxiv.org/abs/1808.06618}{{\tt 1808.06618}}].

\bibitem{Richers:2019grc}
S.~A. Richers, G.~C. McLaughlin, J.~P. Kneller and A.~Vlasenko, \emph{{Neutrino
  Quantum Kinetics in Compact Objects}},
  \href{http://dx.doi.org/10.1103/PhysRevD.99.123014}{\emph{Phys. Rev.} {\bf
  D99} (2019) 123014}, [\href{https://arxiv.org/abs/1903.00022}{{\tt
  1903.00022}}].

\bibitem{Shalgar:2019kzy}
S.~Shalgar and I.~Tamborra, \emph{{On the Occurrence of Crossings Between the
  Angular Distributions of Electron Neutrinos and Antineutrinos in the
  Supernova Core}},
  \href{http://dx.doi.org/10.3847/1538-4357/ab38ba}{\emph{Astrophys. J.} {\bf
  883} (2019) 80}, [\href{https://arxiv.org/abs/1904.07236}{{\tt 1904.07236}}].

\bibitem{Azari:2019jvr}
M.~Delfan~Azari, S.~Yamada, T.~Morinaga, W.~Iwakami, H.~Okawa, H.~Nagakura
  et~al., \emph{{Linear Analysis of Fast-Pairwise Collective Neutrino
  Oscillations in Core-Collapse Supernovae based on the Results of Boltzmann
  Simulations}},
  \href{http://dx.doi.org/10.1103/PhysRevD.99.103011}{\emph{Phys. Rev.} {\bf
  D99} (2019) 103011}, [\href{https://arxiv.org/abs/1902.07467}{{\tt
  1902.07467}}].

\bibitem{Merle:2017dhf}
A.~Merle, \emph{{Sterile Neutrino Dark Matter}}.
\newblock IOP, 2017,
  \href{http://dx.doi.org/10.1088/978-1-6817-4481-0}{10.1088/978-1-6817-4481-0}.

\bibitem{Boyarsky:2018tvu}
A.~Boyarsky, M.~Drewes, T.~Lasserre, S.~Mertens and O.~Ruchayskiy,
  \emph{{Sterile Neutrino Dark Matter}},
  \href{http://dx.doi.org/10.1016/j.ppnp.2018.07.004}{\emph{Prog. Part. Nucl.
  Phys.} {\bf 104} (2019) 1--45}, [\href{https://arxiv.org/abs/1807.07938}{{\tt
  1807.07938}}].

\bibitem{Abazajian:2012ys}
K.~N. Abazajian et~al., \emph{{Light Sterile Neutrinos: A White Paper}},
  \href{https://arxiv.org/abs/1204.5379}{{\tt 1204.5379}}.

\bibitem{Abazajian:2019ejt}
K.~N. Abazajian and A.~Kusenko, \emph{{Hidden Treasures: sterile neutrinos as
  dark matter with miraculous abundance, structure formation for different
  production mechanisms, and a solution to the sigma-8 problem}},
  \href{https://arxiv.org/abs/1907.11696}{{\tt 1907.11696}}.

\bibitem{Xiong:2019nvw}
Z.~Xiong, M.-R. Wu and Y.-Z. Qian, \emph{{Active-sterile Neutrino Oscillations
  in Neutrino-driven Winds: Implications for Nucleosynthesis}},
  \href{https://arxiv.org/abs/1904.09371}{{\tt 1904.09371}}.

\bibitem{Pllumbi:2014saa}
E.~Pllumbi, I.~Tamborra, S.~Wanajo, H.-T. Janka and L.~H{\"u}depohl,
  \emph{{Impact of neutrino flavor oscillations on the neutrino-driven wind
  nucleosynthesis of an electron-capture supernova}},
  \href{http://dx.doi.org/10.1088/0004-637X/808/2/188}{\emph{Astrophys. J.}
  {\bf 808} (2015) 188}, [\href{https://arxiv.org/abs/1406.2596}{{\tt
  1406.2596}}].

\bibitem{Tamborra:2011is}
I.~Tamborra, G.~G. Raffelt, L.~H{\"u}depohl and H.-T. Janka, \emph{{Impact of
  eV-mass sterile neutrinos on neutrino-driven supernova outflows}},
  \href{http://dx.doi.org/10.1088/1475-7516/2012/01/013}{\emph{JCAP} {\bf 1201}
  (2012) 013}, [\href{https://arxiv.org/abs/1110.2104}{{\tt 1110.2104}}].

\bibitem{Wu:2013gxa}
M.-R. Wu, T.~Fischer, L.~Huther, G.~Mart{\'{\i}}nez-Pinedo and Y.-Z. Qian,
  \emph{{Impact of active-sterile neutrino mixing on supernova explosion and
  nucleosynthesis}},
  \href{http://dx.doi.org/10.1103/PhysRevD.89.061303}{\emph{Phys. Rev.} {\bf
  D89} (2014) 061303}, [\href{https://arxiv.org/abs/1305.2382}{{\tt
  1305.2382}}].

\bibitem{Nunokawa:1997ct}
H.~Nunokawa, J.~T. Peltoniemi, A.~Rossi and J.~W.~F. Valle, \emph{{Supernova
  bounds on resonant active sterile neutrino conversions}},
  \href{http://dx.doi.org/10.1103/PhysRevD.56.1704}{\emph{Phys. Rev.} {\bf D56}
  (1997) 1704--1713}, [\href{https://arxiv.org/abs/hep-ph/9702372}{{\tt
  hep-ph/9702372}}].

\bibitem{Ng:2019gch}
K.~C.~Y. Ng, B.~M. Roach, K.~Perez, J.~F. Beacom, S.~Horiuchi, R.~Krivonos
  et~al., \emph{{New Constraints on Sterile Neutrino Dark Matter from $NuSTAR$
  M31 Observations}},
  \href{http://dx.doi.org/10.1103/PhysRevD.99.083005}{\emph{Phys. Rev.} {\bf
  D99} (2019) 083005}, [\href{https://arxiv.org/abs/1901.01262}{{\tt
  1901.01262}}].

\bibitem{Shi:1993ee}
X.~Shi and G.~Sigl, \emph{{A Type II supernovae constraint on electron-neutrino
  - sterile-neutrino mixing}},
  \href{http://dx.doi.org/10.1016/0370-2693(94)90233-X,
  10.1016/0370-2693(94)91232-7}{\emph{Phys. Lett.} {\bf B323} (1994) 360--366},
  [\href{https://arxiv.org/abs/hep-ph/9312247}{{\tt hep-ph/9312247}}].

\bibitem{Hidaka:2007se}
J.~Hidaka and G.~M. Fuller, \emph{{Sterile Neutrino-Enhanced Supernova
  Explosions}}, \href{http://dx.doi.org/10.1103/PhysRevD.76.083516}{\emph{Phys.
  Rev.} {\bf D76} (2007) 083516}, [\href{https://arxiv.org/abs/0706.3886}{{\tt
  0706.3886}}].

\bibitem{Hidaka:2006sg}
J.~Hidaka and G.~M. Fuller, \emph{{Dark matter sterile neutrinos in stellar
  collapse: Alteration of energy/lepton number transport and a mechanism for
  supernova explosion enhancement}},
  \href{http://dx.doi.org/10.1103/PhysRevD.74.125015}{\emph{Phys. Rev.} {\bf
  D74} (2006) 125015}, [\href{https://arxiv.org/abs/astro-ph/0609425}{{\tt
  astro-ph/0609425}}].

\bibitem{Arguelles:2016uwb}
C.~A. Arg{\"{u}}elles, V.~Brdar and J.~Kopp, \emph{{Production of keV Sterile
  Neutrinos in Supernovae: New Constraints and Gamma Ray Observables}},
  \href{http://dx.doi.org/10.1103/PhysRevD.99.043012}{\emph{Phys. Rev.} {\bf
  D99} (2019) 043012}, [\href{https://arxiv.org/abs/1605.00654}{{\tt
  1605.00654}}].

\bibitem{Warren:2016slz}
M.~Warren, G.~J. Mathews, M.~Meixner, J.~Hidaka and T.~Kajino, \emph{{Impact of
  sterile neutrino dark matter on core-collapse supernovae}},
  \href{http://dx.doi.org/10.1142/S0217751X16501372}{\emph{Int. J. Mod. Phys.}
  {\bf A31} (2016) 1650137}, [\href{https://arxiv.org/abs/1603.05503}{{\tt
  1603.05503}}].

\bibitem{Raffelt:2011nc}
G.~G. Raffelt and S.~Zhou, \emph{{Supernova bound on keV-mass sterile neutrinos
  reexamined}}, \href{http://dx.doi.org/10.1103/PhysRevD.83.093014}{\emph{Phys.
  Rev.} {\bf D83} (2011) 093014}, [\href{https://arxiv.org/abs/1102.5124}{{\tt
  1102.5124}}].

\bibitem{Bollig2016}
{MPA Supernova Archive, \url{https://wwwmpa.mpa-garching.mpg.de/ccsnarchive}}.

\bibitem{Keil:2002in}
M.~T. Keil, G.~G. Raffelt and H.-T. Janka, \emph{{Monte Carlo study of
  supernova neutrino spectra formation}},
  \href{http://dx.doi.org/10.1086/375130}{\emph{Astrophys. J.} {\bf 590} (2003)
  971--991}, [\href{https://arxiv.org/abs/astro-ph/0208035}{{\tt
  astro-ph/0208035}}].

\bibitem{Tamborra:2012ac}
I.~Tamborra, B.~M{\"u}ller, L.~H{\"u}depohl, H.-T. Janka and G.~G. Raffelt,
  \emph{{High-resolution supernova neutrino spectra represented by a simple
  fit}}, \href{http://dx.doi.org/10.1103/PhysRevD.86.125031}{\emph{Phys. Rev.}
  {\bf D86} (2012) 125031}, [\href{https://arxiv.org/abs/1211.3920}{{\tt
  1211.3920}}].

\bibitem{Mikheev:1986if}
S.~P. Mikheev and A.~{\relax Yu}. Smirnov, \emph{{Neutrino Oscillations in a
  Variable Density Medium and Neutrino Bursts Due to the Gravitational Collapse
  of Stars}}, {\emph{Sov. Phys. JETP} {\bf 64} (1986) 4--7},
  [\href{https://arxiv.org/abs/0706.0454}{{\tt 0706.0454}}].

\bibitem{1985YaFiz..42.1441M}
S.~P. {Mikheyev} and A.~{\relax Yu}. {Smirnov}, \emph{{Resonance enhancement of
  oscillations in matter and solar neutrino spectroscopy}}, {\emph{Yadernaya
  Fizika} {\bf 42} (1985) 1441--1448}.

\bibitem{1978PhRvD..17.2369W}
L.~{Wolfenstein}, \emph{{Neutrino oscillations in matter}},
  \href{http://dx.doi.org/10.1103/PhysRevD.17.2369}{\emph{Phys. Rev. D} {\bf
  17} (May, 1978) 2369--2374}.

\bibitem{Kim:1987ss}
C.~W. Kim, W.~K. Sze and S.~Nussinov, \emph{{On Neutrino Oscillations and the
  Landau-zener Formula}},
  \href{http://dx.doi.org/10.1103/PhysRevD.35.4014}{\emph{Phys. Rev.} {\bf D35}
  (1987) 4014}.

\bibitem{Parke:1986jy}
S.~J. Parke, \emph{{Nonadiabatic Level Crossing in Resonant Neutrino
  Oscillations}},
  \href{http://dx.doi.org/10.1103/PhysRevLett.57.1275}{\emph{Phys. Rev. Lett.}
  {\bf 57} (1986) 1275--1278}.

\bibitem{Raffelt:1992bs}
G.~G. Raffelt and G.~Sigl, \emph{{Neutrino flavor conversion in a supernova
  core}},
  \href{http://dx.doi.org/10.1016/0927-6505(93)90020-E}{\emph{Astropart. Phys.}
  {\bf 1} (1993) 165--184}, [\href{https://arxiv.org/abs/astro-ph/9209005}{{\tt
  astro-ph/9209005}}].

\bibitem{Abazajian:2001nj}
K.~Abazajian, G.~M. Fuller and M.~Patel, \emph{{Sterile neutrino hot, warm, and
  cold dark matter}},
  \href{http://dx.doi.org/10.1103/PhysRevD.64.023501}{\emph{Phys. Rev.} {\bf
  D64} (2001) 023501}, [\href{https://arxiv.org/abs/astro-ph/0101524}{{\tt
  astro-ph/0101524}}].

\bibitem{Bollig:2017lki}
R.~Bollig, H.-T. Janka, A.~Lohs, G.~Mart{\'{\i}}nez-Pinedo, C.~J. Horowitz and
  T.~Melson, \emph{{Muon Creation in Supernova Matter Facilitates
  Neutrino-driven Explosions}},
  \href{http://dx.doi.org/10.1103/PhysRevLett.119.242702}{\emph{Phys. Rev.
  Lett.} {\bf 119} (2017) 242702},
  [\href{https://arxiv.org/abs/1706.04630}{{\tt 1706.04630}}].

\bibitem{Leitner:2006sp}
T.~Leitner, L.~Alvarez-Ruso and U.~Mosel, \emph{{Neutral current
  neutrino-nucleus interactions at intermediate energies}},
  \href{http://dx.doi.org/10.1103/PhysRevC.74.065502}{\emph{Phys. Rev.} {\bf
  C74} (2006) 065502}, [\href{https://arxiv.org/abs/nucl-th/0606058}{{\tt
  nucl-th/0606058}}].

\bibitem{Hannestad:1997gc}
S.~Hannestad and G.~G. Raffelt, \emph{{Supernova neutrino opacity from
  nucleon-nucleon Bremsstrahlung and related processes}},
  \href{http://dx.doi.org/10.1086/306303}{\emph{Astrophys. J.} {\bf 507} (1998)
  339--352}, [\href{https://arxiv.org/abs/astro-ph/9711132}{{\tt
  astro-ph/9711132}}].

\bibitem{Raffelt:1996wa}
G.~G. Raffelt, \emph{{Stars as laboratories for fundamental physics}}.
\newblock Chicago, USA: Univ. Pr. (1996) 664 p, 1996.

\bibitem{Lamb:1976ac}
D.~Q. Lamb and C.~J. Pethick, \emph{{Effects of neutrino degeneracy in
  supernova models}}, \href{http://dx.doi.org/10.1086/182272}{\emph{Astrophys.
  J.} {\bf 209} (1976) L77--L81}.

\bibitem{Bruenn:1985en}
S.~W. Bruenn, \emph{{Stellar core collapse: Numerical model and infall epoch}},
  \href{http://dx.doi.org/10.1086/191056}{\emph{Astrophys. J. Suppl.} {\bf 58}
  (1985) 771--841}.

\end{thebibliography}\endgroup

\end{document}